\documentclass{aa}
\bibpunct{(}{)}{;}{a}{}{,} 
\pdfoutput=1
\usepackage{hyperref}
\usepackage{graphicx}
\usepackage{txfonts}
\usepackage{amsmath}
\usepackage{float}
\usepackage{physics}
\usepackage{siunitx}
\usepackage{xcolor}

\definecolor{awesome}{rgb}{1.0, 0.13, 0.32}

\begin{document} 



\title{Characterization and dynamics of the peculiar stream Jhelum}
\subtitle{A tentative role for the Sagittarius dwarf galaxy}


 \author{{Hanneke C. Woudenberg} 
          \and
          {Orlin Koop\thanks{The first and second author contributed equally to the work.\\
          e-mail: woudenberg@astro.rug.nl, koop@astro.rug.nl}}
          \and 
          {Eduardo Balbinot}
          \and 
          {Amina Helmi}
          }

   \institute{Kapteyn Astronomical Institute, University of Groningen, Landleven 12, NL-9747 AD Groningen, the Netherlands,}

   \date{Received xxxx; accepted yyyy}

\abstract
   {Stellar streams are a promising tool to study the Milky Way's dark matter subhalo population,
   as interactions with subhalos are expected to leave visible imprints in the streams in the form of  substructure. However, {there may be} other causes of substructure.}
   {Here we {studied} the kinematics and the unusual morphology of the stellar stream Jhelum.}
   {Using a combination of ground-based photometry and \textit{Gaia} EDR3 astrometry, we 
   {characterized}  the morphology of Jhelum. We {combined} this new
   data with radial velocities from the literature to perform orbit integrations of 
   the stream in static Galactic potentials. We also {{carried} out N-body simulations
   in the presence of the Sagittarius dwarf galaxy.}} 
   {The new data reveal a previously unreported tertiary component in the stream, as well 
   as several gaps and a kink-like feature in its narrow component. We find that for a range of 
   realistic Galactic potentials, no 
   single orbit is 
   able to reproduce Jhelum's radial velocity data entirely. A generic property of the orbital solutions is that they share a similar orbital plane to 
   Sagittarius and this leads to repeated encounters with the stream. Using N-body simulations that include a massive Sagittarius, {we {explored} its effect on Jhelum, and} we show that these encounters 
   are able to {qualitatively} reproduce the narrow and broad components in Jhelum, as well as create 
   a tertiary component in some cases. We also find evidence that such encounters 
   can result in an apparent increase in the velocity dispersion of the stream by a factor up to {four}  due to overlapping narrow and broad components.
   }
    {Our findings suggest that the Jhelum stream is even more complex than once though{t;
  however, }its morphology and kinematics can 
  {tentatively} be explained via the interactions with
  Sagittarius. In this scenario, the formation of 
  Jhelum's narrow and broad components occurs naturally, yet some of the smaller gap-like 
  features remain to be explained.}


\maketitle
\section{Introduction} \label{sec:intro}

While orbiting a host galaxy, globular clusters {({GCs})} and dwarf galaxies {({DGs})} experience a
tidal field and may lose mass, producing a stream of stars that approximately traces
their progenitor's orbit \citep{Eyre:2009}. With the advent of large area
photometric surveys, such as the Sloan Digital Sky Survey
\citep[SDSS;][]{York:2000} and{, }more 
recently, the Dark Energy Survey \citep[DES;][]{Shipp_2018}, many such streams have been observed in our Galaxy \citep{Odenkirchen:2001, Newberg:2002, Grillmair:2006, Belokurov:2007} . 

Because of their coherent dynamics, stellar streams are particularly powerful for constraining the mass distribution of the host galaxy. In the case of the Milky Way {(MW)}, they have been used to constrain the properties of its dark matter halo such as the  mass enclosed \citep[e.g.][]{Kupper2015,Pal5GD1MWHalo,constrainMWpotGD1}, its shape \citep[e.g.][]{2001ApJ...551..294I,Law2010,Koposov2010,VeraCiro2013,VasilievTango2021}{,} and its radial density profile  \citep{Gibbons2016}.  

Stellar streams are also promising probes to
detect and study dark matter (DM) subhalos, predicted to orbit the halos of galaxies in large numbers in the case that dark matter is cold \citep{1999ApJ...524L..19M,1999ApJ...522...82K}. Subhalos are expected to produce substructures in the streams such as gaps, overdensities and off-stream features  \citep{Yoon2011Clumpystreams, Erkal2016DMgaps,
Bovy2017, KoppelmanGaps2021}. For example, the off-stream features in the stellar stream GD-1 are thought to have been caused by a substructure of $\sim 10^6 M_\odot$ without a known (visible) counterpart on an orbit close to that of the Sagittarius (Sgr) stream \citep{Bonaca2019GD1spurgap}. Furthermore,
it has been argued that \mbox{GD-1's} overall morphology can constrain the DM subhalo
population \citep{Banik2021GD1subhalos}. Similarly, the perturbed morphology of
the Palomar 5 stream is thought to be partly due to interactions with DM
subhalos \citep{Erkal2017Pal5, Bonaca2020Pal5}.

Dark matter subhalos are not the only objects that can perturb stellar streams. For instance, the gravitational effect of the Large Magellanic Cloud (LMC) needs to be considered to explain the kinematics of
the Orphan stream and the kinematics of the leading arm of the Sgr
stream \citep{VeraCiro2013,Erkal2019Orphan,VasilievTango2021}. 
The peculiar morphology of Palomar 5 is believed to be also in part due to the influence of the Galactic bar \citep{Erkal2017Pal5, Pearson2017, Banik2019}{; also,} spiral arms \citep{Banik2019} and encounters with {giant molecular clouds }are
thought to produce features in streams that may be hard to distinguish
from those induced by the DM subhalos \citep{Amorisco2016, Banik2019}. {Furthermore, chaotic diffusion due to the underlying gravitational potential can also cause distinct stream morphologies \citep{PriceWhelan2016, Yavetz2021} although this depends on the specific region probed by the orbit  \citep{Bonaca2019Jhelum}. A GC progenitor  initially orbiting in a {DM} subhalo may also yield a stellar stream embedded in a diffuse component \citep{Penarrubia2017, Malhan:2019, MalhanAccreted2021}.} 
It is thus clear that,
{before stellar streams can be used to study the DM subhalo population, other possible causes for substructure in stellar streams need to be understood and excluded.}

In this paper, we study the kinematics and morphology of the stellar
stream Jhelum using the recently released DES DR2 and \textit{Gaia} EDR3 data.
Jhelum, spanning about 30 degrees on the {southern} sky, was discovered by
\cite{Shipp_2018} in the multiband optical imaging data from DES. That same year, Jhelum's existence was confirmed in \textit{Gaia} DR2 data
by \cite{Malhan2018}. Subsequent research by \citet{Bonaca2019Jhelum} showed
that the stream consists of a narrow and a broad component.
Recent work suggests that Jhelum
is the remnant of a disrupted {DG} \citep{JiRVJhe, Bonaca2021,
Li2021S5}, to which the Indus stellar stream might also be associated, though
this is under debate \citep{MalhanLMS12021, Li2021S5}. In this
paper, we show that Jhelum's orbital track as delineated by the stream cannot be fitted well in a
static Galactic potential. {We note that Jhelum and Sgr share an orbital plane, and that repeated close encounters could be the cause of the observed stream morphology and substructure.} 

This paper is structured as follows.  Sect.~\ref{sec:datamethod} discusses the
data and the methods used to identify Jhelum members. In Sect.
\ref{sec:orbitfitting}, we present our orbit fitting method and demonstrate that even by varying the characteristic parameters of the {MW} potential, it is difficult  to obtain a reasonable fit to the track delineated by the Jhelum stream. Inspection of the orbits obtained lead us to suggest  that Jhelum experienced an encounter with Sgr. In Sect.~\ref{sec:encounter} we explore this possibility more thoroughly using N-body
simulations, and we demonstrate that such encounters {could} explain the observed morphology of the Jhelum stream.
We end with a general discussion and conclusions in Sect.~\ref{sec:disc_&_conc}.

\section{Data and method} \label{sec:datamethod}

To identify potential members of the Jhelum stream, we cross-    {matched} data
from DES DR2 \citep{DESDR2} and \textit{Gaia} EDR3 \citep{GaiaEDR3} to produce a sample
of high photometric quality stars in a region defined by $-70 < \alpha/\si{deg}
< 30$ and $-60 < \delta/\si{deg} < -40$. We     {selected} ste    {llar-typ}e objects from
DES by filtering in $|$\texttt{WAVG\_SPREAD\_MODEL\_I}$| < 0.01$ and
\texttt{WAVG\_MAG\_PSF\_G} $< 21.5$. All magnitudes     {were} corrected for
extinction using \citet{SFD98} maps, assuming a \citet{Cardelli:1989}
extinction law with $R_{V}=3.1$. This dataset     {was} supplemented with radial velocity
measurements for nine confirmed Jhelum members from \citet{Sheffield2021} and
\citet{JiRVJhe}.

\begin{figure}[t!] \center
    \includegraphics[width=0.9\columnwidth]{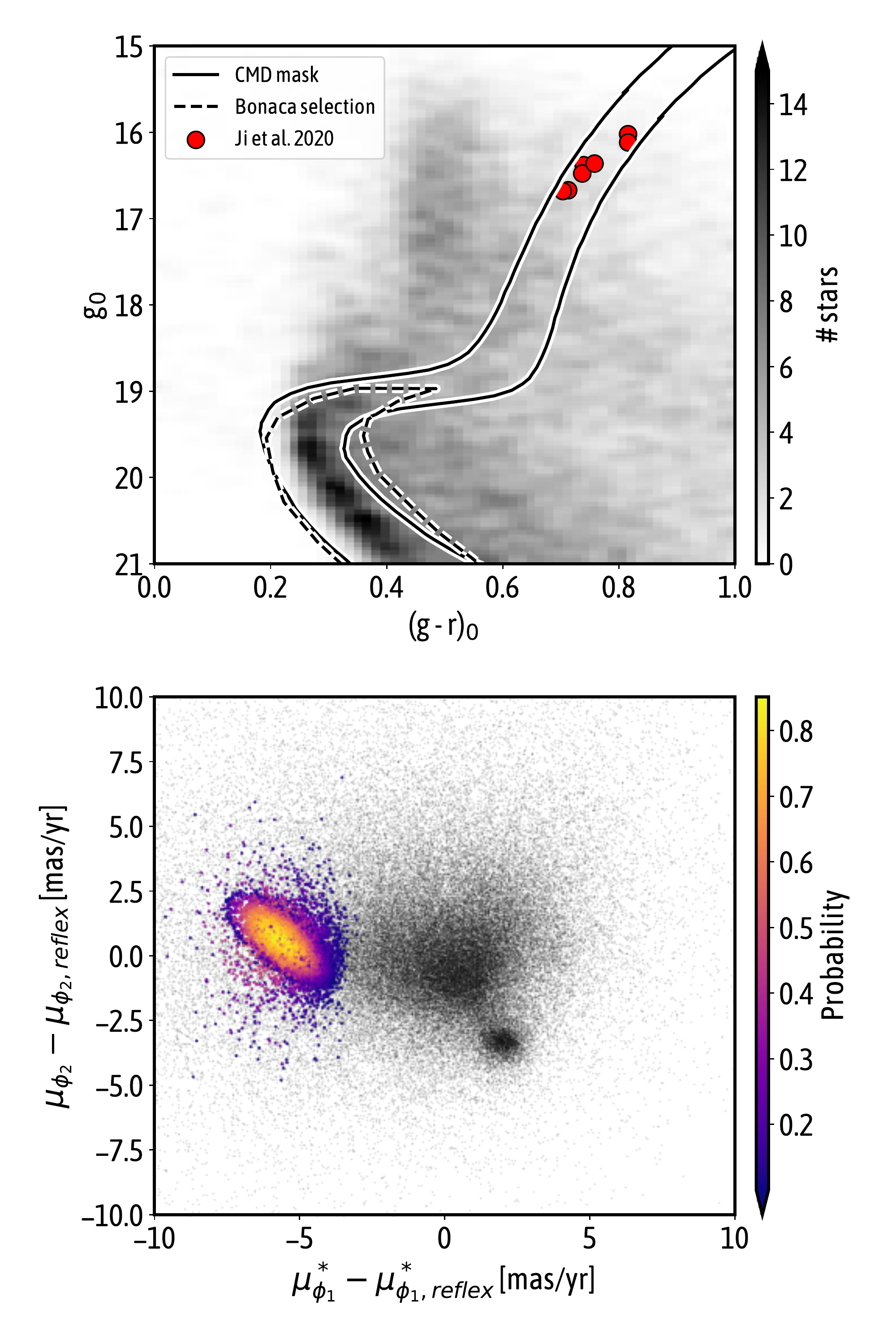}
    \caption{\emph{Top panel:} Hess diagram of stars selected in reflex corrected PM space following 
        \citet{Bonaca2019Jhelum}, and with $|\phi_2| < 2$~deg and \texttt{RUWE} $< 1.4$.  The diagram has been
        smoothed using a 0.4 mag Gaussian kernel. The dashed line shows the
        selection from the same authors, while the solid line shows that adopted
        in this work. Our selection is based on a 12 Gyr, $\textrm{[Fe/H]} = -1.7$
        \texttt{PARSEC} isochrone at a distance of 13 kpc. The red circles show
        confirmed stream members from \citet{JiRVJhe}. \emph{Bottom panel:}
        reflex corrected PM for stars selected according to the CMD mask shown
        above. Jhelum candidates are coloured coded by their membership probability.}
\label{fig:CMD} \end{figure}

Following \cite{Bonaca2019Jhelum} we      {defined} a stream-aligned coordinate system $(\phi_1,
\phi_2, \mu_{\phi_1}^*, \mu_{\phi_2})$\footnote{Throughout this paper, we denote azimuthal proper motions with the shorthand * for convenience. For example, $\mu_{\alpha} \: \cos \delta$
is denoted $\mu_{\alpha}^*$.}.
The pole of this coordinate system is located at 
$ (\alpha, \delta) = (359.1, -141.8)$~deg (this equals the pole found by
\cite{Shipp_2018}, but rotated in $\delta$ by 180 deg), and the origin is
located at $ (\alpha_0, \delta_0) = (359.1, -51.9)$ deg. In this coordinate
system the Jhelum stream is approximately aligned with $\phi_2 = 0$. We     {implemented} this
    coordinate transformation using \texttt{astropy} and{, }to convert from
    Galactocentric to stream coordinates, we     {assumed} that the Galactic Center is
    located at $(\alpha, \delta) = (266.4051, -28.936175)$~deg and that the
    distance from the Sun to the Galactic Center is $R_{\odot} = 8.122 $ kpc
    \citep{Reid2004_galcen, distancegalcen}. The height above the Galactic
    mid-plane is set to $z_{\odot} = 20.8 $ pc \citep{Bennett2019_zsun}.
    Furthermore, we     {adopted} a solar motion of $(U_\odot,V_\odot+V_{\rm
    LSR},W_\odot) = (12.9, 12.2 + 233.4, 7.78)$ \SI{}{km \: s^{-1}}
    \citep{Drimmel2018vsun} and a distance to Jhelum of 13 kpc to correct for
the solar reflex motion, when necessary.

To select regions of the CMD likely to contain Jhelum members, we initially
    {adopted} the proper motion (PM) selection from \cite{Bonaca2019Jhelum} and     {required} that
$|\phi_2|<2$ and \texttt{RUWE} $< 1.4$. This PM selection     {was} done in the reflex corrected
    stream-aligned coordinate system $(\phi_1, \phi_2, \mu_{\phi_1}^* -
    \mu_{\phi_{1, \rm reflex}}^*, \mu_{\phi_2}^* - \mu_{\phi_{2, \rm reflex}}^*)$ following
\cite{Bonaca2019Jhelum}.
We find that a 12 Gyr \textsc{PARSECv1.2S} \citep{Bressan12} isochrone
with $\textrm{[Fe/H]} = -1.7$ at a distance of 13 kpc provides a reasonable
fit to the CMD of the Jhelum stars.  We     {used} the confirmed stream members from \citet{Sheffield2021}
and \citet{JiRVJhe} to anchor the Red Giant Branch (RGB) location, as it is
quite sparse in our DES + \textit{Gaia} dataset. In Fig.~\ref{fig:CMD} we show the CMD mask obtained by selecting a region around the
best-fit isochrone found as described above. 

Using this CMD selection, we     {modelled} the distribution in the reflex motion corrected PM space as a 6
component Gaussian mixture model\footnote{The number of components was
selected using the AIC criteria \citep{AIC}.} \citep{Bovy11}, using the full co-variance matrix (and again imposing that 
$|\phi_2| < 2 \si{\:
deg}$, and \texttt{RUWE} $< 1.4$). 
After this step, we     {evaluated} the Jhelum membership
probability for all stars in our initial sample while relaxing the $|\phi_2|$
selection criterion. In the bottom panel of Fig.~\ref{fig:CMD}, 
we show the PM distribution of all
    stars in our sample, where Jhelum candidate stars are colour coded according to their membership probability (provided this is larger than 10\%).

\subsection{Narrow component member selection} 
\label{sec:datamethod:membsel}

\begin{figure}[t!]
\includegraphics[width=\hsize]{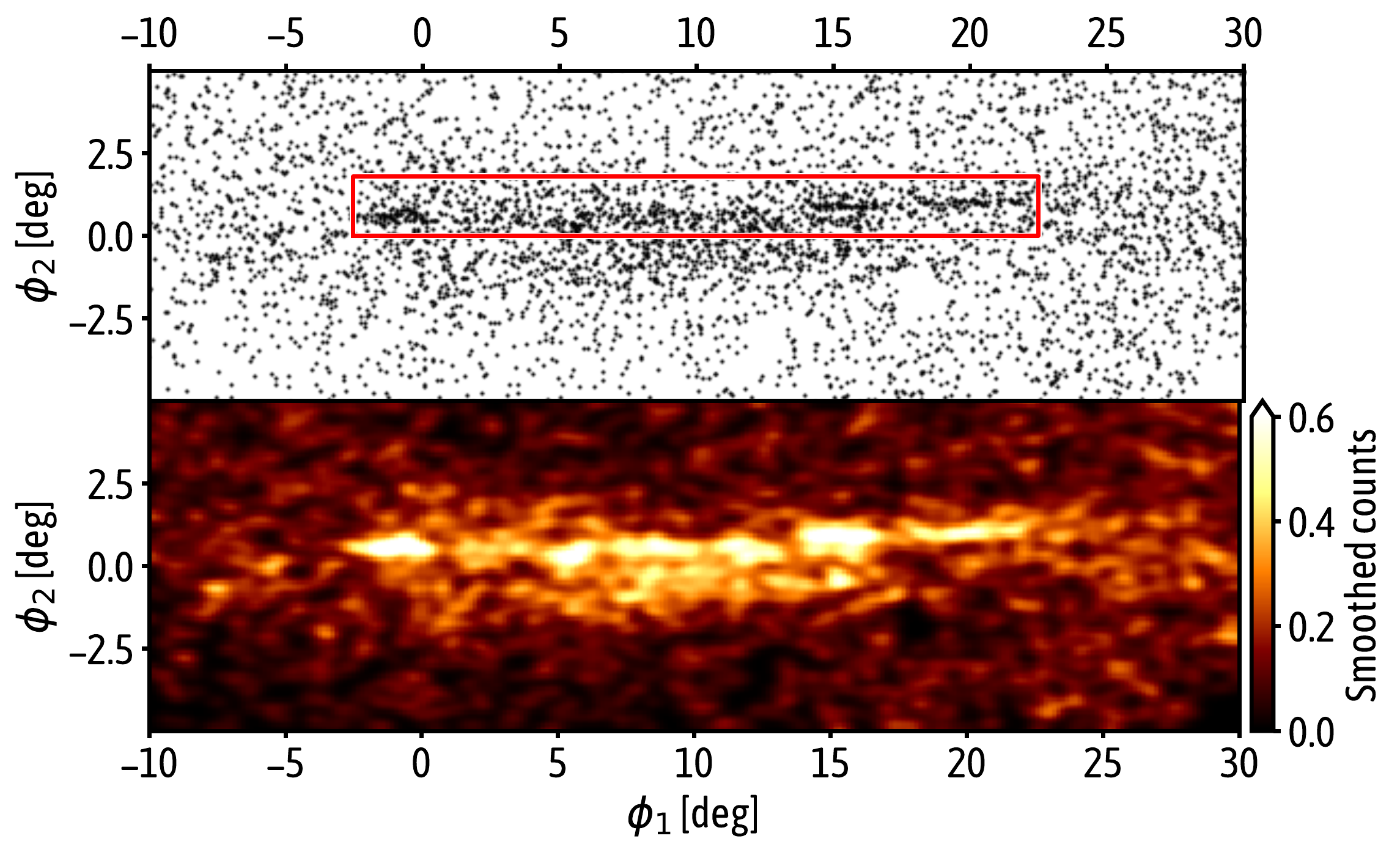}
\caption{Positions in $(\phi_1, \phi_2)$ of stars with a membership probability
    larger than $50 \%$ and parallax $< 0.8$~mas displayed as a scatter plot
    (\emph{top}) and a smoothed density map (\emph{bottom}).  Jhelum clearly
    stands out as an overdensity and has a clumpy morphology with multiple gaps. Possible narrow-component members are contained within the red region 
    in the top panel. Below this region, the extended broad component stands
    out, while above it, between $\phi_1 \approx$ 0 -- 5 degrees, a thus far 
    unreported third component can be seen.} \label{fig:members_NC}
\end{figure}

In Fig.~\ref{fig:members_NC}, we     {show} stars with a membership probability higher than 50\% and with a parallax $< 0.8$~mas.  Especially the smoothed density map plotted in the lower panel of
this figure clearly shows that Jhelum
consists of a narrow component centred around $\phi_1 \approx 0.5$~deg and
an extended broader component beneath it.  Moreover, we find a third,
yet unreported component above the narrow one around $\phi_1 \approx$ 0 -- 5
degrees. The main narrow component has a clumpy morphology with several tentative gapsand 
some overdensities, most notably at $\phi_1 =  -1$ and 16~deg. Finally,
we note that there seems to be a discontinuity in the narrow component at
$\phi_1 \approx 14$~deg. We conclude that Jhelum has a very complex morphology, with many of the features seen in \textit{Gaia} DR2 \citep{Bonaca2019Jhelum} now more clearly apparent 
thanks to the better astrometric quality of \textit{Gaia} EDR3.

\cite{Bonaca2019Jhelum} fitted the median location of the narrow component as a function of $\phi_1$ as 
\begin{equation} 
\phi_2 = 0.000546 \phi_1^2 - 0.00217 \phi_1 + 0.583
\label{eq:median} 
\end{equation}
where $\phi_1$ and $\phi_2$ are in degrees. Therefore, we select as tentative
narrow-component members those stars with membership
probability higher than 50\% and parallax < 0.8 mas that lie within $^{+1}_{-0.6}$~degrees of the
median within the range $-2.5 \leq \phi_1 \leq 22.5$~deg.

\begin{figure}
    \includegraphics[width=\hsize]{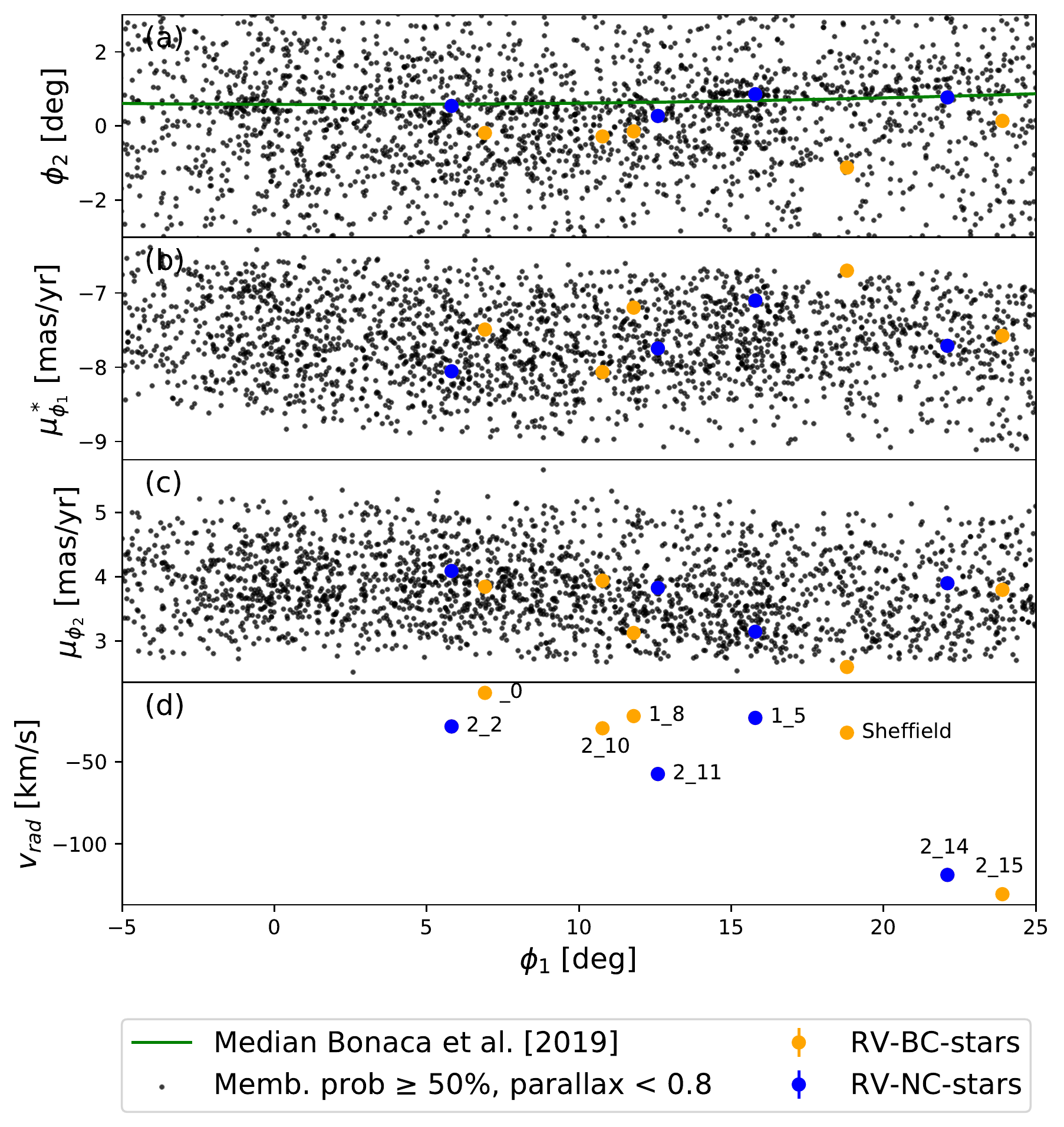} \caption{Stars
        with a membership probability larger than $50 \%$, parallax $< 0.8$~mas and
        $5 \leq \phi_1/ \si{deg} \leq 25$, $-3 \leq \phi_2/ \si{deg}  \leq 3$
        are shown as black dots in panel (a) on the sky $(\phi_1, \phi_2)$, in (b) in $(\phi_1,
        \mu_{\phi_1}^*)$ and in panel (c) $(\phi_1, \mu_{\phi_2})$. The RV-NC-stars are
        overplotted in blue, the RV-BC-stars in orange. Panel (d) shows the radial velocities of the 
        RV-NC-stars and RV-BC-stars, where the labels
        indicate the notation used for the \texttt{JhelumX\_Y} stars. The measured
        uncertainties in position, PM and radial velocity for the
        RV-NC-stars and RV-BC-stars are plotted as errorbars, but are too small
        to be seen. The green curve in the top panel shows the median track (Eq.~\ref{eq:median}) from \cite{Bonaca2019Jhelum}.  \label{fig:RVBCNCstars}} \end{figure}

Of the nine confirmed Jhelum stream members with radial velocity measurements from 
\citet{JiRVJhe,Sheffield2021}, we spatially     {selected} the four stars with $\phi_2 > 0.2 $~deg
as possible narrow component (NC) members.
These are the stars \texttt{Jhelum2\_2}, \texttt{Jhelum1\_5},
\texttt{Jhelum2\_11} and \texttt{Jhelum2\_14} from \cite{JiRVJhe}. Throughout
this work, we refer to these stars as RV-NC-stars. The other five stars,
\texttt{Jhelum\_0}, \texttt{Jhelum1\_8}, \texttt{Jhelum2\_10} and
\texttt{Jhelum2\_15} from \cite{JiRVJhe} and the APOGEE Jhelum giant from
\cite{Sheffield2021}, are assumed to belong to the broad component (BC) and
are referred to as RV-BC-stars. The RV-NC-stars and RV-BC-stars are shown in Fig.~\ref{fig:RVBCNCstars} in blue and orange respectively.  

\subsection{Characterization of the Jhelum stream track} \label{sec:binningprocedure}

To characterize the Jhelum stream and its track in PM,  the positions and
PM of the possible narrow-component members (i.e. inside the box shown in Fig.~\ref{fig:members_NC})     {were} binned in $\phi_1$ with a binsize of $5$~deg and a stepsize
of $2.5$~deg. In
each $\phi_1$-bin we     {fitted} a Gaussian distribution with a constant background $B$, given by
\begin{equation} 
    f(x) = B + A \exp{- \frac{(x - \mu_x)^2}{2 \sigma_x^2}}    {,}
\label{eq:gaussianbinmodel} 
\end{equation} 
for each observable $x$,     {that is to say} $\phi_2, \mu^*_{\phi_1}$, or $\mu_{\phi_2}$, and 
where $A$ is the relative normalization of the Gaussian component, $\mu_x$ is
the mean and $\sigma_x$ is the dispersion. We used the minimization implementation from \texttt{scipy.optimize.minimize} to determine the various parameters in Eq.~\ref{eq:gaussianbinmodel}, all of which are allowed to vary from bin to bin. 

To avoid that particularly high density regions end up with unrealistically low uncertainties, after fitting we     {checked} that in each $\phi_1$-bin, $\sigma_{\phi_2}$ is greater than 
the average $\langle \sigma_{\phi_2}\rangle$ determined using all $\phi_1$-bins. If this is not the case, we     {assigned} it this average. This floor in $\sigma_{\phi_2}$ can be seen as a way to reduce the impact of the clumpiness in Jhelum's narrow component when characterizing the stream track.  

\begin{figure}[t!]
    \includegraphics[width=\hsize]{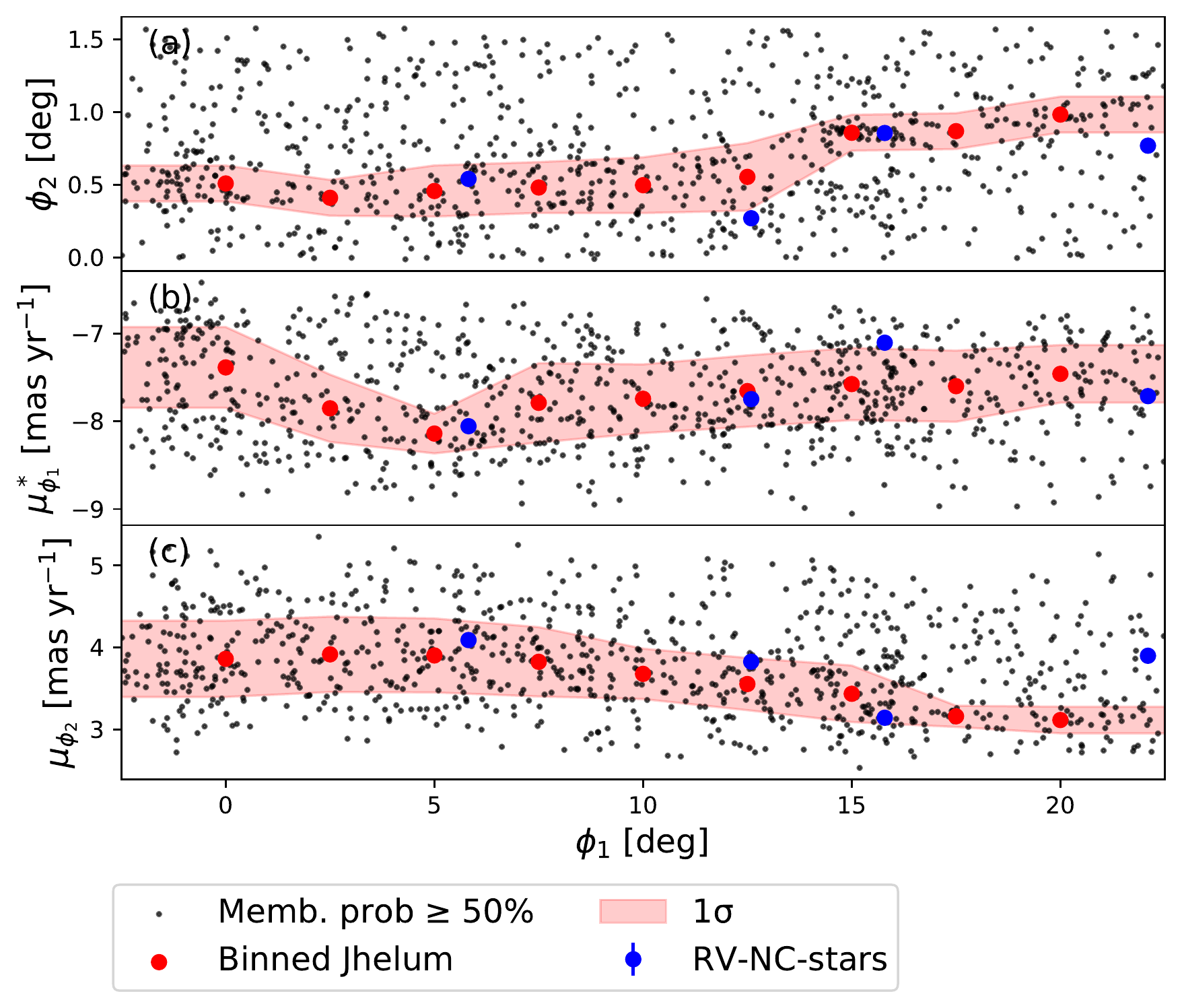}
    \caption{
        Distribution of high probability narrow component members (>50\%)
        in (a) $(\phi_1, \phi_2)$, (b)
        $(\phi_1, \mu_{\phi_1}^*)$ and (c) $(\phi_1, \mu_{\phi_2})$. The
        resulting binned tracks and their $1 \sigma$ uncertainty ranges are
        indicated in red. The RV-NC-stars are overplotted in blue, together
        with their measurement uncertainties (which are too small to be seen). Most 
        RV-NC-stars follow the tracks roughly within the $1\sigma$ range, except for the RV-NC-star \texttt{Jhelum2\_14} located at $\phi_1 \approx 22$~deg.
\label{fig:binningresultshaded}} \end{figure}

The resulting stream track is shown in Fig.~\ref{fig:binningresultshaded},
where the shaded region denotes the extent defined by 1$\sigma$ around the mean estimated as just described.
    {We note} that in the top panel there is a jump in the stream track on the sky around $\phi_1 \approx 14$~deg (see also Fig.~\ref{fig:members_NC}), i.e.
for $\phi_1 < 14$ degrees, the track is approximately constant at $\phi_2
 \approx 0.5$~deg, while for  $\phi_1 > 14$~deg, it is approximately constant but now at $\phi_2 \approx 0.9$~deg.

\section{Orbit fitting} 
\label{sec:orbitfitting}

We now aim to determine if there are
orbits which, when integrated in a     {MW} potential, approximately follow the track defined by the Jhelum stream. Although we do not expect a stream to exactly follow a single orbit \citep[see][for a detailed discussion]{Sanders2013}, it turns out that for Jhelum, this is not a bad approximation (see Appendix~\ref{app:simple-Nbody}).

To establish the characteristics of such an orbit, we     {defined} a log-likelihood which yields the probability that the single orbit fits each datapoint $i$, given the uncertainties in the data and the internal dispersion of the stream, namely 
\begin{multline}
\ln(L) = \frac{1}{N} \sum_{i}^N \ln(L_i) \\ = \frac{1}{N} \sum_{i}^N \left[- \ln \left(\prod_{j=1}^{4} (2 \pi)^{1/2} \sigma_{i,j} \right) - \frac{1}{2} \sum_{j=1}^{4} \left( \frac{x_{i,j}^d - x_{i,j}^m}{\sigma_{i,j}} \right)^2 \right]
\label{eq:negloglikelihoodperdatum_main}
\end{multline}
where $j$ represents the subspace $\delta$, $\mu_{\alpha}^*$, $\mu_{\delta}$ or $v_{rad}$. The superscripts $d$ and $m$ denote the data and model respectively.
The datapoints consist of the means and dispersions obtained using the fitting procedure described in the previous section, as well as the measurements for the individual RV-NC stars.  For the RV-NC stars' astrometry, we     {added} in quadrature to their measurement uncertainties the average values of $\langle \sigma \rangle$ for $\phi_2, \mu^*_{\phi_1}$, and $\mu_{\phi_2}$ as derived in the previous section. This is done to account for the observed dispersion in the stream, which is not reflected in the measurement uncertainties of the individual stars.  

To estimate the uncertainties in the RV-NC stars' radial velocities we proceed as follows. We not only     {considered} the measurement error, but also that the stream might have an internal velocity dispersion, so
$\sigma_{v_{rad}} = \sqrt{\sigma_{\rm meas}^2 +
\sigma_{\rm nuis}^2 }$. The nuisance
parameter, $\sigma_{\rm nuis}$, has two components: a random 
$\sigma_{\rm rand}$, and a systematic $\sigma_{\rm syst}$, which are added in quadrature. The systematic
component     {was} determined by taking the average of the difference in radial
velocity measurements by AAT and MIKE of the 8 \cite{JiRVJhe} stars \citep[see
Table 1 in][]{JiRVJhe}, while leaving out \texttt{Jhelum2\_2} because it is
probably a binary star. This gives $\sigma_{\rm syst} = 1 \si{\: km \: s^{-1}}$, which is consistent with the median velocity offset found by \cite{JiRVJhe}. The
random component     {was} determined by considering three pairs of stars that lie close
together on the sky, \texttt{Jhelum\_0} and \texttt{Jhelum2\_2} around $\phi_1
\approx 6$~deg, \texttt{Jhelum1\_8} and \texttt{Jhelum2\_10} around $\phi_1
\approx 11$~deg, and \texttt{Jhelum2\_14} and \texttt{Jhelum2\_15} around
$\phi_1 \approx 22$~deg, and by measuring their radial velocity difference. We then set the random velocity to be the mean of these differences
which gives $\sigma_{\rm rand} =
6.6 \si{\: km \: s^{-1}}$. Finally, $\sigma_{\rm nuis} = 6.7 \si{\: km \: s^{-1}}$.

    {We note} that this value of $\sigma_{\rm nuis}$ is likely an upper limit to the internal dispersion of the
stream. In practise, a large value allows a more generous range of the free
parameters that are fitted, as can be seen from Eq.~\ref{eq:negloglikelihoodperdatum_main}. This is why we also
    {considered} a more realistic $2 \si{\: km \: s^{-1}}$ for $\sigma_{\rm nuis}$ in what
follows \citep{Kuzma2015, constrainMWpotGD1, Gialluca2021}. 

To start the orbit integrations, 
we     {assumed} a fixed right ascension of $\alpha = 330$~deg, which corresponds to $(\phi_1, \phi_2) \approx (18.7, 0.8)$~deg in the stream-aligned coordinate system. For the other coordinates, we set a flat prior given by 
\begin{equation}
P(\vec{\theta}) = \left\{
    \begin{array}{ll}
        1 & \text{if} \left\{
            \begin{array}{lllll}
                -50.8^{\circ} < \delta < -46.8^{\circ}  \\
                10 < d < 15 \: \text{kpc} \\
                5 < \mu_{\alpha}^* < 8 \: \si{mas.yr^{-1}} \\
                -7 < \mu_{\alpha}^* < -3 \: \si{mas.yr^{-1}} \\
                -125 < v_{rad} < 60 \: \si{km.s^{-1}} 
            \end{array}
            \right.  \\
        0 & \text{otherwise.} 
    \end{array}
\right.
\label{eq:Jhelumprior}
\end{equation}
As we have no
reliable information on the distances to the stars, we     {left} this as a free
parameter within the range of the prior. For the potential, we first     {assumed} a standard     {MW} mass model (see Sect.~\ref{sec:results:orbitfit}), and we then     {explored} what happens when its characteristic parameters are allowed to change (see Sect.~\ref{sec:results:varying}) using the package \texttt{AGAMA} \citep{AGAMA}.

\subsection{Best-fit orbit in a standard Milky Way potential} \label{sec:results:orbitfit}

We    {modelled} the    {MW} gravitational potential as a three-component
system consisting of a bulge, disk, and dark halo, following \citet{galaZenodo}. The bulge    {was} modelled as a Hernquist sphere \citep{Hernquist1990} with a mass of $4\times 10^9 M_{\odot}$ and scale length $c_b = 1$~kpc. The disk    {was} modelled as a Miyamoto-Nagai potential \citep{MiyamotoNagai} with $M_{disk} = 5.5 \times 10^{10} M_{\odot}$, scale length $a_d = 3$~kpc, and scale height $b_d=280$~pc \citep{Bovy2014Potential}. The dark halo follows a generalized Navarro-Frenk-White (NFW) potential \citep{NFWprofile}, which can be made spherical, flattened or triaxial by changing the values of the axis ratios. We set the scale radius $r_s = 15.62$~kpc, $M_{halo} = 0.7 \times 10^{12} M_{\odot}$, and the halo potential to be flattened with a minor-to-major axis ratio $q_z=0.95$. 

Using this potential, we    {performed} a Markov Chain Monte
Carlo (MCMC) with the package \texttt{emcee} \citep{emcee} to find the orbit 
(model) parameters that fit the data best, where we use 80 walkers and 1000
steps. As $\alpha$ is fixed, there are thus five free parameters to be explored
by the MCMC algorithm, ($\delta$, $d$, $\mu_{\alpha}^*$, $\mu_{\delta}$,
$v_{rad}$). An initial guess that    {was} used in the MCMC    {was} found using
\texttt{scipy.optimize}. 

\begin{figure}[htb!]
\includegraphics[width=\hsize]{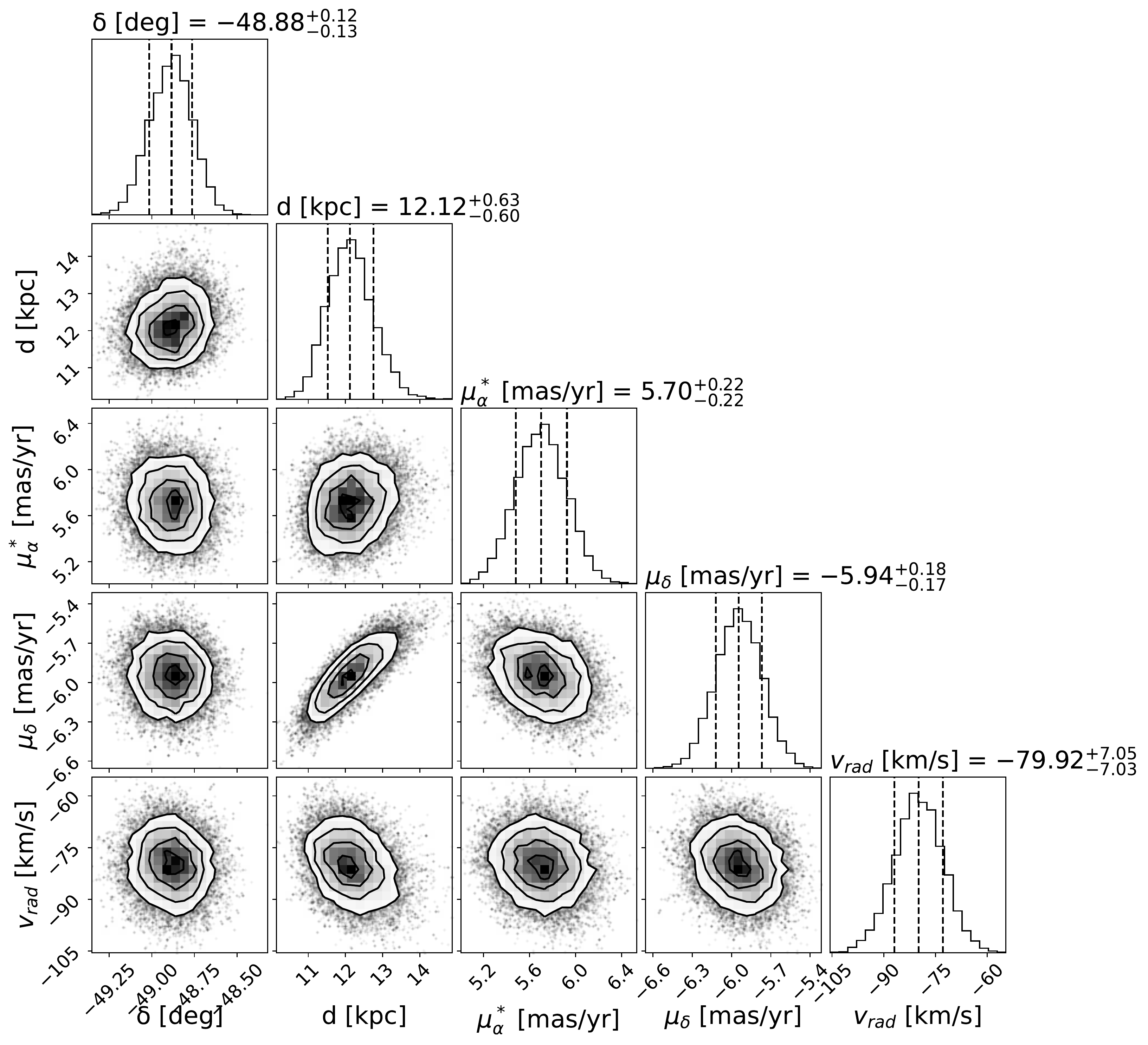}
\caption{Posterior parameter distribution of the initial conditions of the best-fit orbit of Jhelum for $\sigma_{\rm nuis} = 6.7 \si{\: km.s^{-1}}$ in the default MW potential.    {We note} the (expected) strong degeneracy between the distance and $\mu_{\delta}$.
\label{fig:cornersphericalhalofit}}
\end{figure}

The MCMC chains converged regardless of the choice of $\sigma_{\rm nuis}.$ The
resulting posterior parameter distribution obtained for $\sigma_{\rm nuis} = 6.7
\si{\:km.s^{-1}}$  is shown in Fig.~\ref{fig:cornersphericalhalofit}. The posterior parameter distribution for  $\sigma_{\rm nuis}
= 2 \si{\:km.s^{-1}}$ is similar but tighter, especially in radial velocity.
Fig.~\ref{fig:cornersphericalhalofit} reveals that there is a strong degeneracy between the distance and $\mu_{\delta}$, and weaker 
degeneracies between the distance and $v_{rad}$ and between
$\mu_{\delta}$ and $\mu_{\alpha}^*$. The best-fit orbit for $\sigma_{\rm nuis} =
6.7 \si{\:km.s^{-1}}$ has an apocenter of \SI{23.6}{kpc} and pericenter of
\SI{8.1}{kpc} (similar values are found for $\sigma_{\rm nuis} = 2
\si{\:km.s^{-1}}$), where the apo- and pericenter are calculated by averaging them over an integration time of 10 Gyr. Interestingly, although \cite{Bonaca2019Jhelum} did not take radial velocity information into
account, they    {obtain} a similar best-fit orbit, with an 
apocenter of \SI{24}{kpc} and a pericenter of \SI{8}{kpc}.

\begin{figure}[t!]
\includegraphics[width=\hsize]{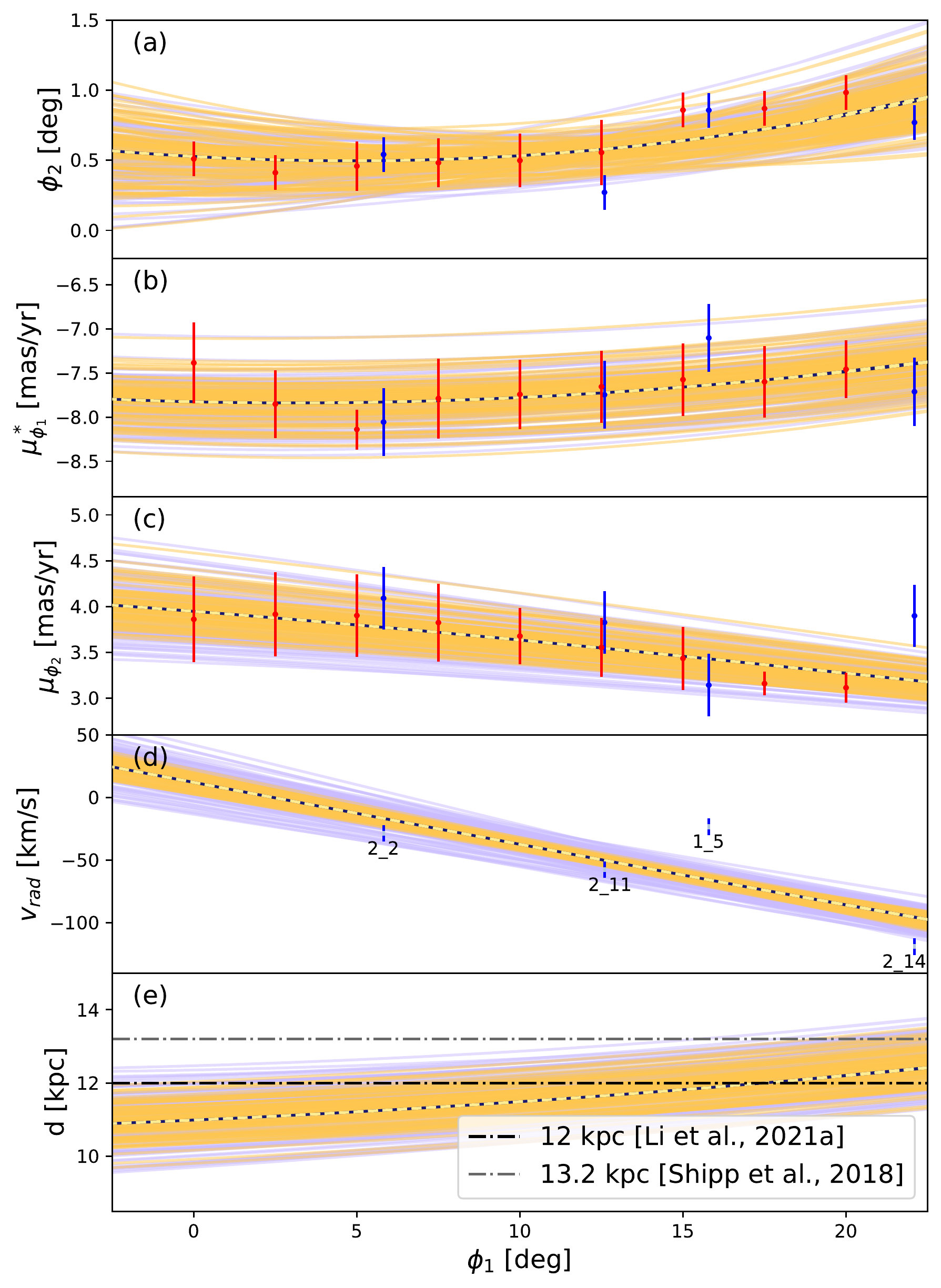}
\caption{Best-fit single orbit of Jhelum in the default    {MW} potential for
    $\sigma_{\rm nuis} = 6.7 \si{\: km.s^{-1}}$ (dark blue line) and
    $\sigma_{\rm nuis} = 2 \si{\: km.s^{-1}}$ (dashed light orange line). 100 orbits    {were} sampled
    randomly from the MCMC chains for the higher and lower values of $\sigma_{\rm nuis}$, and are shown in light indigo and in orange respectively. The binned
    estimates and their $1\sigma$ uncertainties are plotted in red, while the individual RV-NC-stars (along with their names)
    are plotted in blue for $\sigma_{\rm nuis} = 6.7 \si{\: km.s^{-1}}$ (and light-blue for its smaller value). 
    Though the orbits follow the binned track and RV-NC-stars well within the
    error bars in position and    {PM}, in radial velocity no orbits go through all of the four RV-NC stars:
    \texttt{Jhelum2\_14} and especially \texttt{Jhelum1\_5} lie far outside the
    sampled range in radial velocity. The distance to the stream is somewhat
    low in comparison to the estimate by \cite{Shipp_2018}, but consistent with the
    estimate by \cite{Li2021S5}.}
\label{fig:Jhelumsphericalbestfitsampled_phi1phi2}
\end{figure}

The best-fit orbits follow well the positions and
   {PMs} of the stream, see panel (a), (b) and (c) of  Fig.~\ref{fig:Jhelumsphericalbestfitsampled_phi1phi2}, though \texttt{Jhelum2\_14} deviates by about $2 \sigma$ in $\mu_{\phi_2}$ from the best-fit orbit. In fact, \texttt{Jhelum2\_14} was already seen to deviate from the stream track in Fig.~\ref{fig:binningresultshaded}. 
 The
distance distribution is consistent with the recent estimate by \cite{Li2021S5},
but slightly below the estimate by \cite{Shipp_2018}, see panel (d).    {Figure}.~\ref{fig:Jhelumsphericalbestfitsampled_phi1phi2} also shows a sample of 100
orbits, randomly selected from the  posterior distribution of initial conditions. No single orbit is capable of going through all the radial velocity measurements of the 
RV-NC stars, with \texttt{Jhelum2\_14} and \texttt{Jhelum1\_5} being the most
deviant ones. This is the case regardless of the $\sigma_{\rm nuis}$ adopted.

\subsection{Varying gravitational potential parameters} 
\label{sec:results:varying}

\begin{figure*}[hbt!]
\includegraphics[width=\hsize]{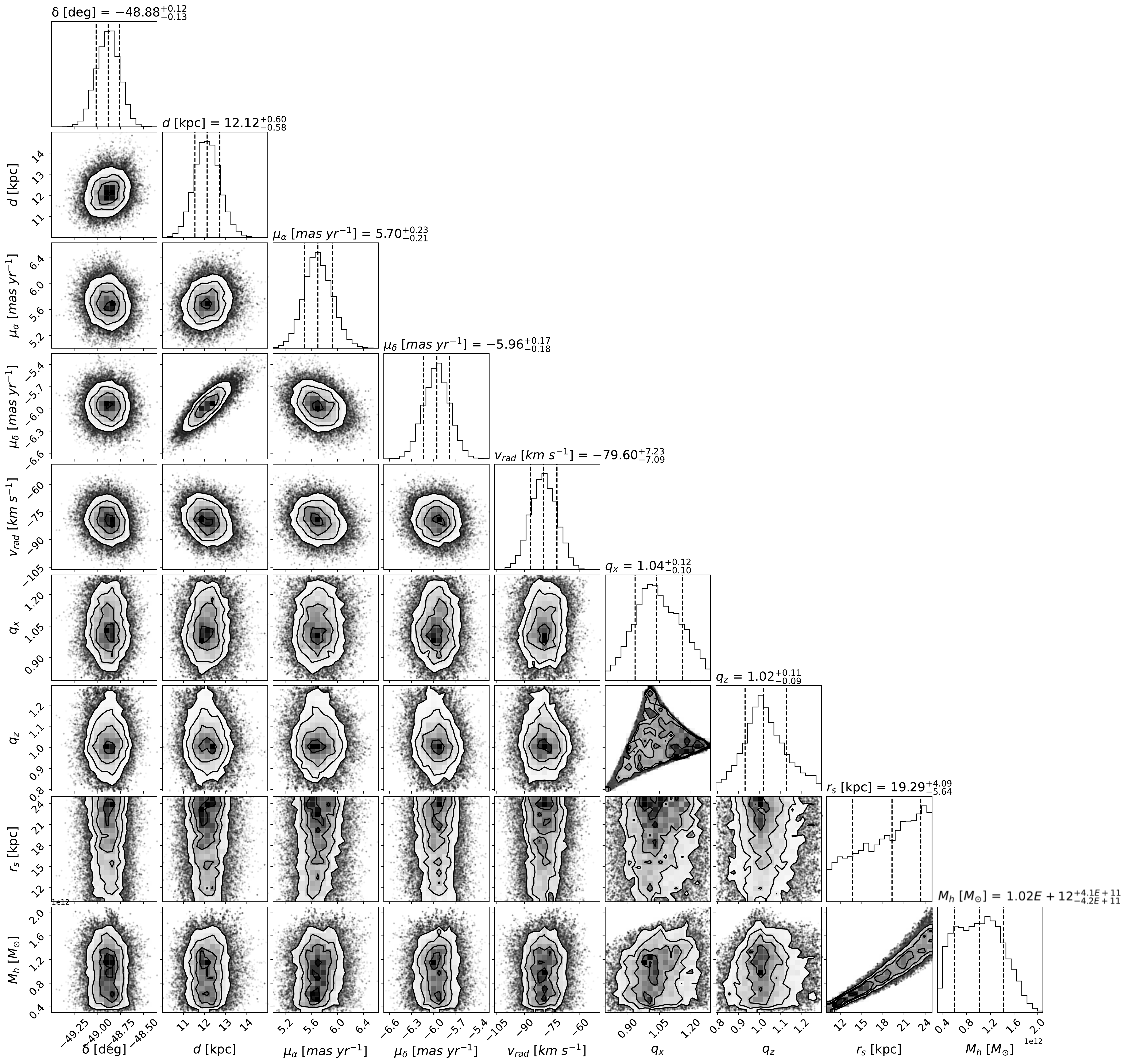}
\caption{Posterior parameter distribution of the initial conditions of Jhelum's best-fit orbit and of the parameters of the Galactic halo potential, for $\sigma_{\rm nuis} = 6.7 \si{\: km/s}$.    {We note} that here, the bulge and disk were kept fixed and set to be those of the default MW potential. The posterior parameter distribution of the initial conditions resembles the one of the default MW potential shown in Fig.~\ref{fig:cornersphericalhalofit}. There is a strong degeneracy between $M_{h}$ and $r_s$, as expected, and is a result of the circular velocity constraint. The correlation between $q_x$ and $q_z$ is set by the requirement that the halo density be positive.}
\label{fig:triaxialcornerplot}
\end{figure*}

We    {explored} if by varying parameters of the Galactic gravitational potential we are able to fit the data on the Jhelum stream better. 
We    {kept} the bulge and disk the same as in the default MW potential, and let the halo be triaxial, where the axis ratios denoted as $q_x = a/b$, $q_z = c/b$,
the scale radius $r_s$ and its mass $M_{halo}$ are all allowed to vary.  We    {restricted} ourselves to potentials that give rise to a correct velocity at the position of the Sun within 5\% assuming $V_{\rm circ}(R_{\odot}) = 233 \si{\: km.s^{-1}}$ \citep{Reid2014, McMillan17, Hayes2018, Eilers2019, Mroz2019}\footnote{   {We note} that the default MW potential model has $V_{\rm circ}(R_{\odot}) = 230.8 \si{\: km.s^{-1}}$ and thus lies within this range.}. We also    {required} that the density of the NFW halo component is positive over the extent of the MW ($-300 < x, y, z < 300 \si{\: kpc}$). These two constraints limit the extent of the parameter space that the MCMC is allowed to explore. As the parameters $q_x$ and $q_z$ but also $r_s$ and $M_{halo}$ are correlated in non-trivial ways, we    {chose} somewhat generous flat priors as a starting point, namely $0.7 < q_x < 1.4$, $0.7< q_z < 1.4$, $3 \cdot 10^{11} < M_{halo} < 3 \cdot 10^{12} M_{\odot}$ and $10 < r_s < 25$~kpc. We then    {checked} 
that both constraints    {were} satisfied by the MCMC chains. 

Together with the parameters associated to the orbital initial conditions 
($\delta$, $d$, $\mu_{\alpha}^*$, $\mu_{\delta}$, $v_{rad}$), we thus have a total of 9 free
parameters to be determined by the MCMC. As the parameter space to be explored by the MCMC has increased, we    {used} 80 walkers and 2000 steps. We only    {considered} the $\sigma_{\rm nuis} = 6.7 \si{\: km.s^{-1}}$ case as this serves effectively an upper limit on the allowed range of the free parameters. The resulting posterior parameter distribution is shown in Fig.~\ref{fig:triaxialcornerplot}, where that of the orbital initial conditions resembles that obtained for the default MW potential fit and shown in Fig.~\ref{fig:cornersphericalhalofit}. 
   {Figure}.~\ref{fig:triaxialcornerplot} shows that there is a strong degeneracy between $M_{halo}$ and the scale radius $r_s$, which is partly induced by the circular velocity constraint. $q_x$ and $q_z$ are also correlated in a non-trivial manner, which is largely determined by the positive density constraint. In comparison to the default halo potential model, a larger scale radius, $r_s = 19 \si{\: kpc}$, in combination with a larger halo mass, $M_h = 1.0 \cdot 10^{12} \: M_{\odot}$, seem to be preferred, as well as an almost spherical halo, with $q_x = 1.04$ and $q_z = 1.02$\footnote{The values of the characteristic parameters could be somewhat biased because an orbit-fitting technique is used to match the stream, rather than for example, an N-body simulation, see \citet{Sanders2013}.}. 

\begin{figure}[t!]
\includegraphics[width=\hsize]{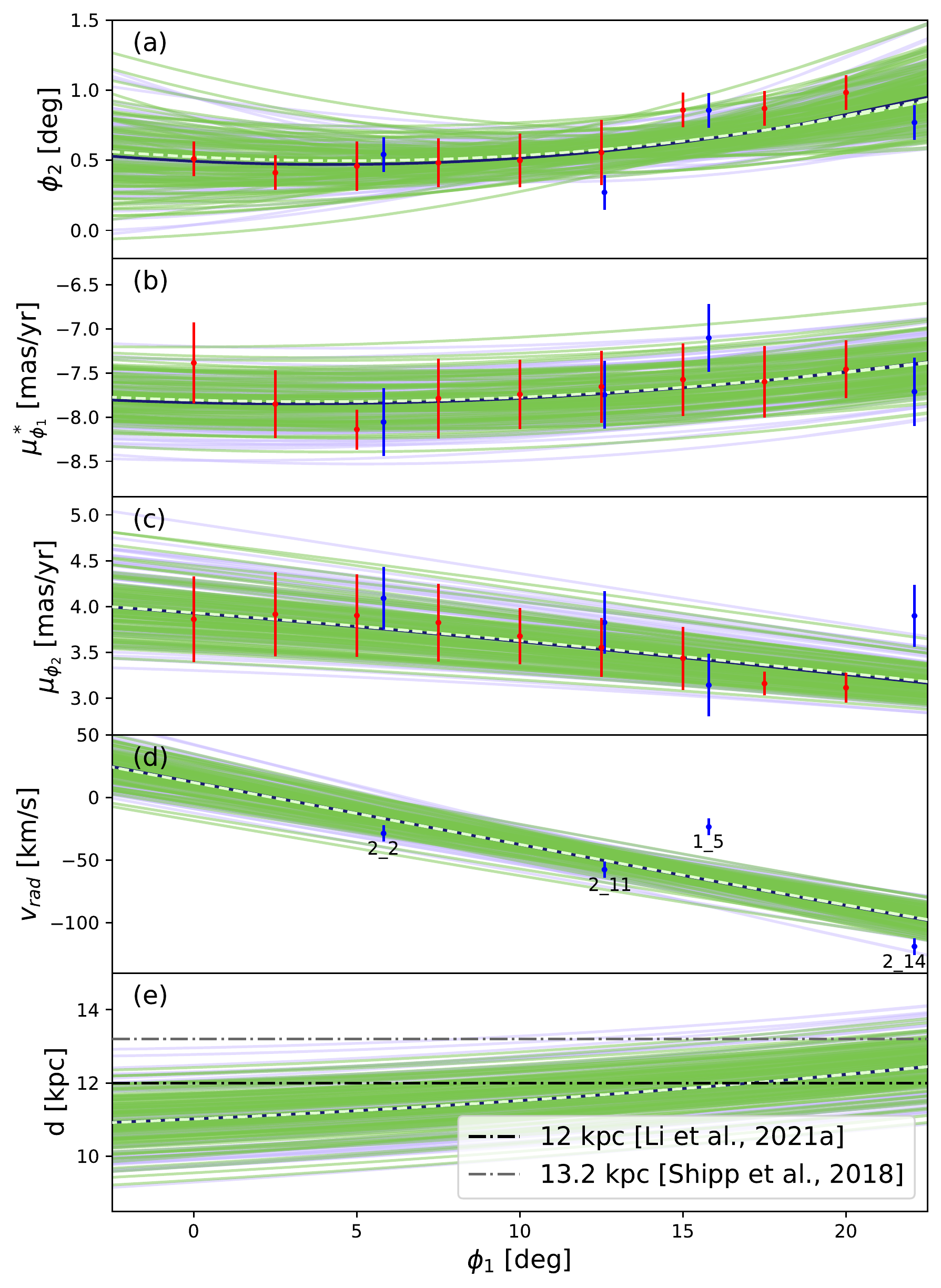}
\caption{Best-fit single orbit of Jhelum for the default MW potential (dark blue line) and the MW potential with triaxial halo (dashed light green line) for $\sigma_{\rm nuis} = 6.7 \si{\: km.s^{-1}}$. 100 orbits sampled randomly from the MCMC chains are shown using the same colour as the corresponding best-fit orbits.    {We note} that allowing parameters of the Galactic potential to vary given the constraints does not yield significantly different orbits. 
\label{fig:overlayJhelumorbits}}
\end{figure}

The best-fit single orbit, as well as 100 orbits sampled randomly from the MCMC chains are shown in Fig.~\ref{fig:overlayJhelumorbits}. For comparison, the best-fit orbit in the default MW potential is shown as well. From 
Fig.~\ref{fig:overlayJhelumorbits} it is clear that allowing the parameters of the potential to vary does not result in a significantly different best-fit orbit (although the data is generally fitted better as with the additional degrees of freedom there are more high-likelihood values). In fact, the two best-fit orbits closely resemble each other in all subspaces. We find again that the best-fit orbits follow the positions and    {PMs} well (with \texttt{Jhelum2\_14} still deviating by more than 2$\sigma$ from the best-fit orbit in $\mu_{\phi_2}$). On the other hand, in radial velocity \texttt{Jhelum2\_14} and \texttt{Jhelum2\_2} lie consistently $10 - 20 \si{\:km.s^{-1}}$ below the best-fit orbits and on the edge of the sampled range, while \texttt{Jhelum1\_5} lies far above it, $\sim 50 \si{\:  km.s^{-1}}$. In conclusion, there is no potential that fits the data perfectly, and in particular the radial velocities cannot be matched, independently of the values of the characteristic parameters or shape of the Galactic potential used.

\subsection{The effect of perturbers: an encounter with Sagittarius} \label{sec:results:encounter}

Jhelum is a peculiar stream with multiple  components showing over- and underdensities (see Fig.~\ref{fig:members_NC}). It is perhaps in hindsight not so surprising that no single orbit in a realistic Galactic 
gravitational potential can be found that matches the track of the narrow component of the stream well.

\begin{table}
\caption{Orbital properties of the LMC and M54 (which is used to trace Sgr's orbit). 
\label{tab:LMCM54props}}
\centering    
\begin{tabular}{ccc} 
\hline\hline 
 & LMC & Sgr (M 54) \\
\hline  
$\alpha$ (deg) & 80.84561 \tablefootmark{a}  & 283.764 \tablefootmark{c}    \\
$\delta$ (deg) & -69.78267 \tablefootmark{a} & -30.480 \tablefootmark{c}   \\
$d$ (\si{kpc}) & 50.6 \tablefootmark{a}      & 26.5 \tablefootmark{d}   \\
$\mu_{\alpha}^*$ (\si{mas.yr^{-1}})& 1.910 \tablefootmark{b} & -2.680 \tablefootmark{c}  \\
$\mu_{\delta}$ (\si{mas.yr^{-1}}) & 0.229 \tablefootmark{b} & -1.387 \tablefootmark{c}   \\
$v_{rad}$ (\si{km.s^{-1}})      & 262.2 \tablefootmark{a} & 141.3 \tablefootmark{d}  \\
\hline   
\end{tabular}
\tablefoot{ \tablefoottext{a}{ \cite{LMCprops}}, \tablefoottext{b}{\cite{LMCpm}}, \tablefoottext{c}{\cite{VBSgr}}, and \tablefoottext{d}{\cite{HSgr}}.}
\end{table}

\begin{figure*}[hbt!]
\includegraphics[width=\hsize]{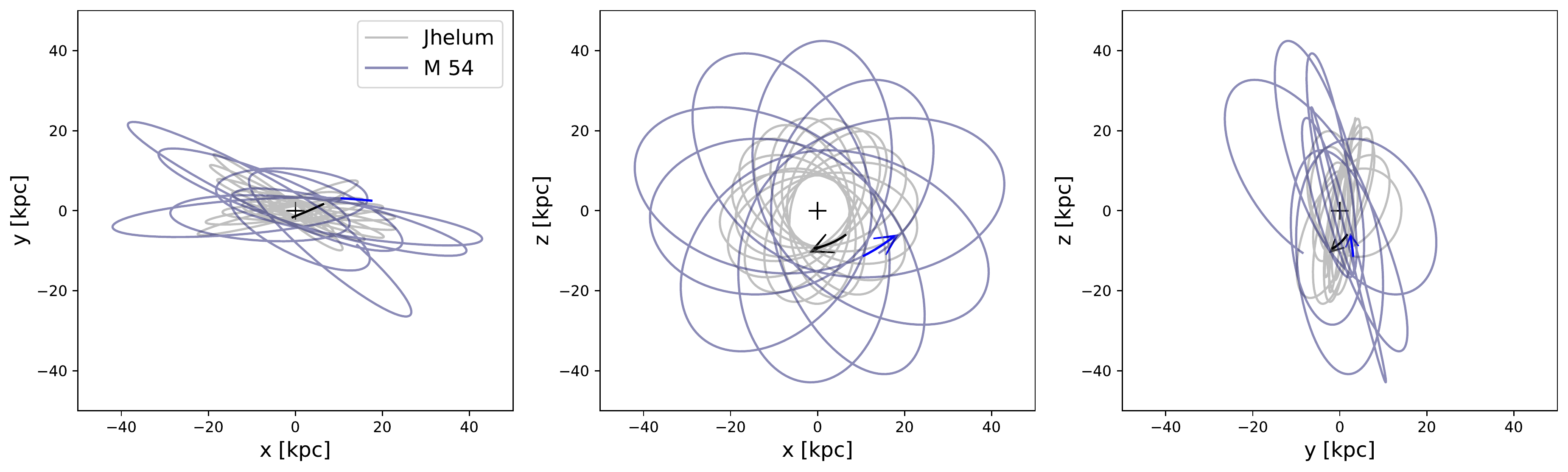}
\caption{Best-fit orbit of Jhelum in the default MW potential (for
    $\sigma_{\rm nuis} = 6.7 \si{\: km.s^{-1}}$) plotted together with the orbit of
    Sgr (using the coordinates of M 54, see Table~\ref{tab:LMCM54props}), both integrated backwards in time for 6 Gyr. The arrows
    indicate the direction in which Jhelum and Sgr are currently moving. It is
    clear that Jhelum and Sgr share approximately the same orbital plane, which
    results in repeated close encounters between the two.   { We note }that Jhelum is currently passing through pericenter, which is the reason for the high observed velocity
gradient seen in Fig.~\ref{fig:overlayJhelumorbits}.
\label{fig:JhelumM54passage}}
\end{figure*}

In search for an explanation, we    {considered} the possibility that
Jhelum has been perturbed while orbiting the    {MW}. We    {focused} in particular on the influence of the two heaviest dwarf satellites galaxies of the    {MW}, namely  the LMC and Sgr. 

To check whether these objects might have come close enough to Jhelum to perturb it, we
   {integrated}, using the default    {MW} potential, the orbit of the LMC {including dynamical friction},
that of Sgr and Jhelum's best-fit single orbits (for both values of  $\sigma_{\rm nuis}$) separately 6 Gyr back in time. The initial conditions for the orbit integrations of Sgr and the LMC are listed in
Table~\ref{tab:LMCM54props}. 
We then    {calculated} the distance at the closest passage between
the centre of mass of Jhelum (which we    {considered} to be located on the best-fit orbit) and these objects. In the case of the LMC, and being on its first infall, the closest passage with Jhelum is at the present time at a relative distance of $44~$\si{kpc}, and although this might have an effect on its dynamics \citep[see e.g. ][]{ShippLMC2021}, such an interaction would not explain its disturbed morphology { as substructure would need more time to form}.
{ To further assess the effects of the LMC in conjunction with Sgr, we    {investigated} their combined influence in Appendix \ref{app:LMCinfluence}. We find again that the LMC's influence only becomes significant recently, which is in line with the findings of \cite{ShippLMC2021}.}

On the other hand, Jhelum and Sgr passed each other closely multiple
times over the past 6 Gyr, with distances at closest passage  ranging from  $3 - 6$~kpc.     {Figure.}~\ref{fig:JhelumM54passage} shows the orbit of Jhelum and Sgr both integrated for \SI{6}{Gyr} back in time and reveals that these objects seem to
share a similar orbital plane, which explains why they have
repeated close encounters. This is also the case for the    {MW} potential fitted in Sect.~\ref{sec:results:varying}.
{The periodic character of these interactions suggests that Sgr might have had  a considerable influence on Jhelum, and that this could perhaps explain its orbit and the observed peculiar stream morphology.}

 In the case of
the LMC, and being on its first infall, the closest passage with
Jhelum is at the present time at a relative distance of 44 kpc and
although it might have an effect on its dynamics, such an interaction would not explain its disturbed morphology

\section{Study of the Sgr-Jhelum interaction}\label{sec:encounter}

  {In this section, we further investigate
the possibility that Jhelum was perturbed by Sgr. To this end, we performed N-body simulations of the interaction between the progenitor of Jhelum and Sgr while they orbit the Galaxy.}

\subsection{Choice of initial conditions}
\label{sec:orbits_IC_Nbody}
To explore a set of reasonable orbital initial conditions for the progenitor of Jhelum we   {created} a bundle of possible orbits around one Jhelum member, \texttt{Jhelum2\_2}, which we set to represent the present-day position of the progenitor. We used $\alpha$ and
$\delta$ from \texttt{Jhelum2\_2} and sampled $\mu_{\alpha}^*$ and
$\mu_{\delta}$ from a Gaussian distribution with $\sigma$ given by the error estimated in Sect.~\ref{sec:binningprocedure}. For the distance we   {sampled} from a uniform distribution between \SI{10.6}{kpc} and \SI{14.6}{kpc}, corresponding to the value reported for the Jhelum member
in \cite{Sheffield2021} and compatible with photometric distance estimates
to the stream. We   {sampled} the initial radial velocity from a Gaussian distribution with mean
and variance from the observed RV-NC stars, excluding \texttt{Jhelum2\_14} and
\texttt{Jhelum2\_15} which deviate from the seemingly linear velocity trend (see Fig. \ref{fig:RVBCNCstars}).

\begin{table*}
\caption{{Sampling parameters for the initial conditions of the orbit integrations of Jhelum and the parameters of the four selected orbits used for the N-body experiments integrated with GADGET-4.   {For these integrations,} $\alpha$ and
$\delta$ were kept  {at a} fixed   {value}).}} 
\label{tab:Simprops}
\centering
\begin{tabular}{ccc|cccc} 
\hline \hline 
 &   {Jhelum} & $\sigma$ & Sim 1 & Sim 2 & Sim 3 & Sim 4 \\
\hline 
$\alpha$ (h:m:s) & 23:18:34.74 \tablefootmark{a}  &  --  & -- & -- & -- & -- \\
$\delta$ (d:m:s) & 52:02:10.2 \tablefootmark{a}  &  --  & -- & -- & -- & -- \\
$d$ (\si{kpc}) & 12.6$\pm2.0$ \tablefootmark{b}    & -- & 10.7 & 11.1 & 11.6 & 10.7\\
$\mu_{\alpha}^*$ (\si{mas.yr^{-1}})& 7.46\tablefootmark{a} &  0.51 & 7.74 & 7.45 & 7.31 & 7.50\\
$\mu_{\delta}$ (\si{mas.yr^{-1}}) & -5.10\tablefootmark{a}  & 0.64   & -5.04  & -4.90 & -4.71 & -5.04\\
$v_{rad}$ (\si{km.s^{-1}}) & -28.8\tablefootmark{c}  &  13.8 & -25.7 &  -39.3 & -24.8 & -25.4\\
\hline 
\end{tabular}
\tablefoot{\tablefoottext{a}{\cite{GaiaDR22018}}
\tablefoottext{b}{\cite{Sheffield2021}} \tablefoottext{c}{\cite{JiRVJhe}}{. E}rrors for the Jhelum {PMs} were found by binning \textit{Gaia} EDR3 data of
  {Jhelum} and the error for the radial velocity was taken equal to the ensemble
variance of the selected Jhelum members in \cite{JiRVJhe}. The properties for
  {Jhelum} were taken from the star labelled \texttt{Jhelum2\_2} in \cite{JiRVJhe}.}
\end{table*}

We then   {integrated} the orbits generated this way {backwards in time}, in the same Galactic potential as 
described in Sect.~\ref{sec:results:orbitfit}{. We now also}   {included} the potential associated to Sgr. 
For Sgr, we   {used} the orbital initial conditions listed in Table \ref{tab:LMCM54props}. {As a starting point}, we   {fixed} the mass of Sgr initially to $3\times10^9M_{\odot}$ and 
  {assumed that} it follows a Plummer profile with a scale radius of $\sim 1$~kpc, based on a rescaling of the model in \cite{Laporte2018} after converting to a Plummer sphere with a similar concentration parameter \citep[as in Appendix B of][]{KoppelmanGaps2021}. 

Before deciding which of the generated initial conditions {for Jhelum} may
be most interesting for the N-body simulations, we   {explored} the orbits further. Firstly we   {considered}
only those orbits that reasonably reproduce the track of Jhelum on the sky. The interpolated orbit (for each initial condition generated as described above)   {was} compared to the 
observed stream track using the quantity $D^2$, defined as 
\begin{equation}
   D^2 = \sum_{i}^{N} (\phi_{2,\mathrm{pol}}(\phi_{1,i})-\phi_{2,{\mathrm{orb}}}(\phi_{1,i}))^2,\label{eq:D2}
\end{equation}
where $\phi_{2,\mathrm{pol}}$ is the polynomial from Eq.~\eqref{eq:median}
and $N=40$ is the number of points where the  polynomial is compared to the orbit. 

We consider orbits with $D^2 < 0.2$ to follow the stream track well (top 1.73\% of the 50000 generated initial conditions).   {We note} that since $D^2$ is calculated using on-sky angular distances, there is a bias towards preferring orbits with a larger distance, as we see below. 
  {During} our orbit integrations we   {kept} track of 
the minimal relative distance and relative velocity to Sgr. We focus on interactions with Sgr that occurred more than 2~Gyr ago, as more recent ones would likely not produce easily visible perturbations in the stream {as the timespan is too short, and also because Sgr would likely be lighter and hence have a smaller impact \citep[e.g.][]{LastBreathSgr}.} From this sample, we   {selected} four orbits that come closer than {6}~kpc to Sgr and {which, as we show below}, produce interesting features in {the Jhelum stream}.

\subsection{Simulation setup} \label{sec:encounter:simulation}
While previous works   {conclude} that Jhelum most likely had a   {DG}
progenitor \citep[e.g.][]{Li2021S5,Shipp2019pmcomp,Bonaca2019Jhelum,JiRVJhe},   {we choose to take a (loose) GC progenitor}.   {We do this because the} narrow component of Jhelum has a width comparable to that of other cold streams \citep[e.g.][]{Shipp_2018} and our
goal   {is} to fit the narrow component while perturbing the stream to 
produce the   {observed features}.
The progenitor follows a Plummer sphere of scale radius \SI{9}{pc} with a
mass of $5\times10^3M_{\odot}$, and consists of $5\times 10^3$ particles, and was generated self-consistently using \texttt{AGAMA} \citep{AGAMA}. The   {chosen} scale radius is typical for observed GCs
\citep{deBoer2019}, while the mass is somewhat lower 
\citep{Baumgardt2018}, to allow for a cold enough stream with a fully dissolved progenitor after 6 Gyr of evolution.

{To model the interaction between Jhelum and Sagittarius, one would perhaps like to resort to full N-body simulations representing Jhelum and its stream,   {Sgr,} and the underlying Galactic potential. This is however prohibitive, as this would require these systems to be represented by particles with comparable masses to prevent artificial heating in the (newly formed) stream \citep[see][]{Banik2021}. Given the mass of Jhelum and the desired resolution in its stream, this implies $m_p \sim 1 M_{\odot}$. This would require $10^{12}$ particles for the   {MW} halo (and $\sim 10^9$ particles for Sgr). Since the use of currently computationally feasible simulations (with $m_p \sim 10^3 - 10^5 M_{\odot}$) would incorrectly model the gravitational potential of the   {MW}, we prefer to use the analytic description presented in   {Sect.}~\ref{sec:results:orbitfit}.}

{ A full N-body realization of Sgr would have the  advantage of accounting for mass loss{,} which may have an effect on the strength of the interactions with Jhelum. Therefore, to still model the mass loss of Sgr, we first   {ran} an N-body simulation of Sgr evolving on its own in the background   {MW} potential. We then   {used} this simulation to calibrate a time-dependent analytic Plummer potential of Sgr as this evolves. This   {was} subsequently employed in a   {forwards} N-body simulation of Jhelum to model Sgr as one particle with a softening scale and mass determined by the Plummer parameters at each point in time.} 

{For the N-body simulations of Sgr, we   {used} \texttt{AGAMA} to generate $10^5$ particles according to a self-consistent distribution function belonging to the Plummer sphere from   {Sect.}~\ref{sec:orbits_IC_Nbody}, and let it dynamically relax for $\sim6$~\si{Gyr}.
We then   {carried} out the N-body simulation of Sgr in the   {MW} potential integrating   {forwards} in time for $\sim 6$~Gyr. We   {used} GADGET-4 \citep{GADGET-4} with \texttt{SELFGRAVITY} enabled, and   {included} the background   {MW} potential using the \texttt{EXTERNALGRAVITY} flag. At each snapshot, we
  {fitted} an analytic Plummer potential given by}
\begin{equation}
    M_{\mathrm{Plummer}}(<r) = M\dfrac{r^3}{(r^2+r_s^2)^{3/2}}\label{eq:plummermassenc}
\end{equation}
{to the distribution of particles still bound to the progenitor. This results in a tabulated mass and scale radius for Sgr through time. We fit this evolution in time with smooth polynomials. This then   {allowed} us to represent closely the behaviour of a live Sgr evolving in a Galactic potential.} { We show the evolution of the parameters of our Sgr model through time in Appendix \ref{app:sgrparams}.} 

We   {performed} our N-body simulations {of the Jhelum stream} with the same configuration for GADGET-4 \citep{GADGET-4}   {as before}. We   {considered} two particle types: one for Jhelum stars and one for {the single} particle representing Sgr. For the Sgr particle, we   {adopted} a softening parameter equal to its {fitted} scale radius, which in practise corresponds to the Plummer sphere representing Sgr. 
The softening length for the Jhelum particles was set to 2~pc \citep[determined using the methodology of][]{Athanassoula2000,Villalobos2008}.
{We   {proceeded} in two steps. We first   {integrated} Jhelum starting from the initial conditions selected in   {Sect}.~\ref{sec:orbits_IC_Nbody}   {backwards} with the mass varying Sgr particle. Then we   {replaced} the Jhelum particle with a relaxed progenitor and   {integrated}   {forwards} again. In all our simulations}
we allow a maximum timestep of \SI{0.49}{Myr}, with a default
\texttt{ErrTolIntAccuracy} of 0.012.

To study the specifics about the stream-formation process, we   {monitored} the history of mass loss.
We   {defined} the centre of mass of the progenitor by taking the mean position of the 200
particles with the highest binding energy (identified initially, i.e. in the relaxed progenitor), and 
keeping track of those that remain within the tidal radius $r_t$ at all times. The tidal radius   {was} calculated by using the constraint that the equipotential
surface between the background potential ($\rho_{\mathrm{bg}}$) and the
progenitor ($\rho_{\mathrm{prog}}$) occurs at the tidal radius,   {that is to say}
$\rho_{\mathrm{prog}}(r_t)=3\rho_{\mathrm{bg}}(r_{\mathrm{peri}})$, where
$r_{\mathrm{peri}}$ is the pericentric distance of the progenitor. In our case, this results in $r_t = 18$~pc. 
Whenever a particle crosses the tidal radius for the last time in the
simulation, we say it has become unbound.   {We record the following pericenter
passage of the progenitor as the release time of the particle}.

\subsection{Tracking and characterization of the interactions}\label{sec:encounter:interactions}

Close interactions between Sgr and Jhelum may result in velocity kicks to the stream members. Such kicks can generally be described using the impulse approximation. If the impact parameter is $b$ (i.e. the distance between the perturber and the point of closest impact on the stream) and the stream is aligned with the $y$-direction, the maximal total velocity kick $\Delta v$ imparted at a given point is 
\begin{equation}
    \Delta v_{\rm max} = \dfrac{2GMb}{w(b^2+r_s^2)},
\end{equation}
where $M$ is the mass, and $r_s$ the scale radius of the perturber. Here $w=(w_{\parallel}^2+w_{\perp}^2)^{1/2}$, with
$w_{\parallel}$ the relative $y$-velocity of the perturber along the stream and
$w_{\perp}=(w_x^2+w_z^2)^{1/2}$, with $w_{x,z}$ the respective components of the
velocity of the perturber. We refer the reader to \cite{Carlberg2013} and \cite{ErkalBelokurov2015} for general expressions for the velocity impulse 
experienced by stream particles
due to the passage of a perturber. 

To monitor the amplitude of the velocity kicks and the location of the closest encounters, we first   {identified} the snapshots where the distance between Sgr and the Jhelum progenitor   {is} smallest. Then we   {searched} in the neighbouring (five preceding and following) snapshots and   {identified} the   {snapshot where} the largest number of particles   {are} close to Sgr. The point at which this occurs is used to label the time of the interaction and to compute the velocity kick experienced by those particles. 

\subsection{Results}\label{sec:encounter:results}

\begin{figure}
\includegraphics[width=.95\linewidth]{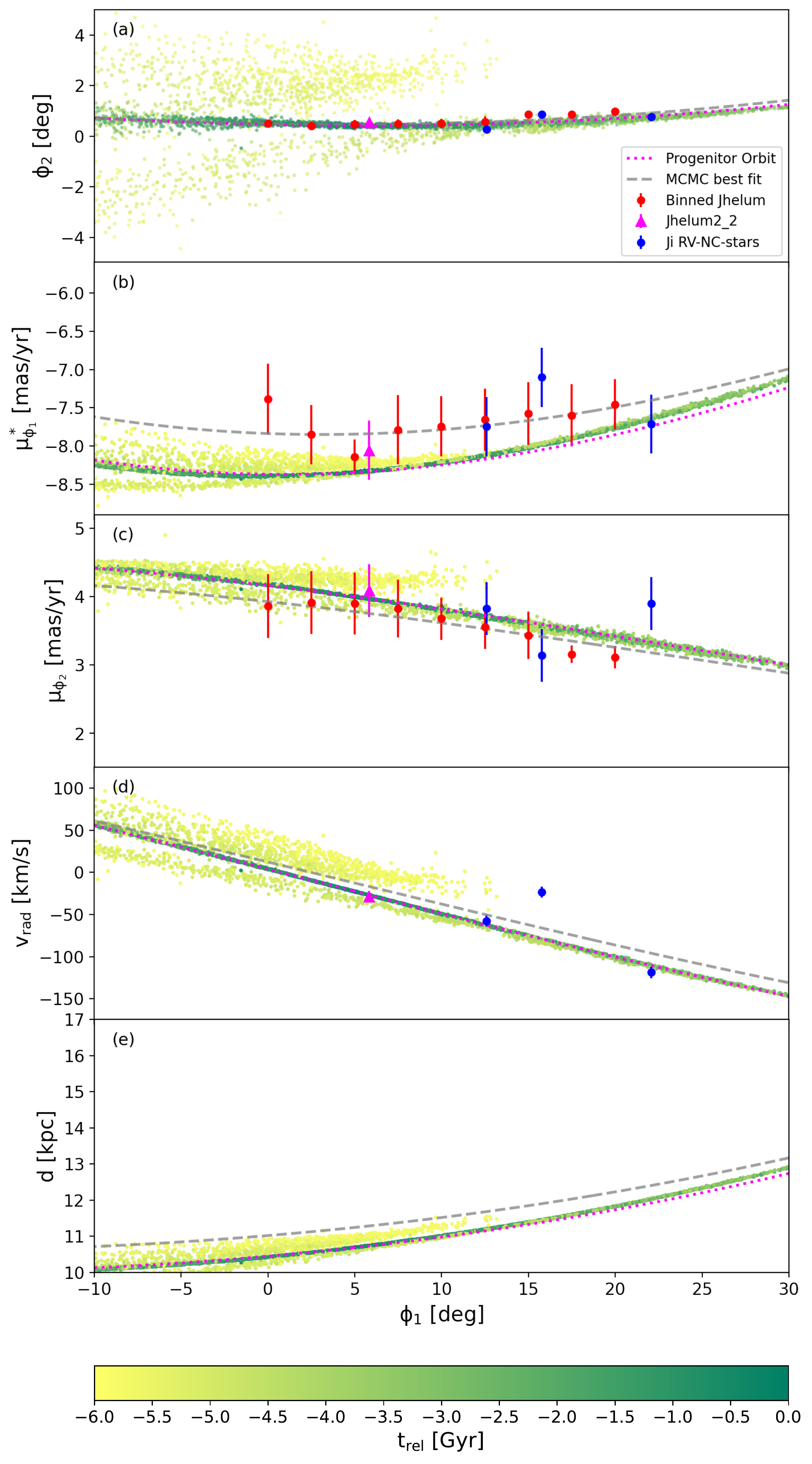}
\caption{Simulation 1 of the Jhelum stream after interacting with the   {Sgr} dwarf spheroidal. The individual star particles from Jhelum are   {colour}-coded according to
    the lookback time at which they became unbound.
    Panel (a) shows the stream in $\phi_1$-$\phi_2$ frame, while panels (b) and (c) show the
    (non-reflex corrected) PM in the $\phi_1$ and $\phi_2$
    directions respectively. Panel (d) shows the radial velocity, and Panel (e)
    the distance. The red datapoints correspond to the binned
    Jhelum data from Sect.~\ref{sec:binningprocedure}, and the blue datapoints
    correspond to stars from \cite{JiRVJhe} that are likely members of the
    narrow component of Jhelum. The magenta triangle corresponds to 
    \texttt{Jhelum2\_2}, the datapoint which in our simulations marks the present-day location
    of the progenitor. The magenta dotted line shows the trajectory of the   {centre}
    of mass of the Jhelum progenitor, and the grey dashed line shows the result
    from the MCMC best-fit with $\sigma_{\mathrm{nuis}}=6.7~\si{km.s^{-1}}$ from 
    Sect.~\ref{sec:results:orbitfit}. 
Panel (a) shows clear secondary and tertiary components, which are also apparent in {the other panels}. These additional components are made up of particles that were released early on in the formation of the stream.
\label{fig:Sim1}}
\end{figure}

\begin{figure}[t!]
\includegraphics[width=.95\linewidth]{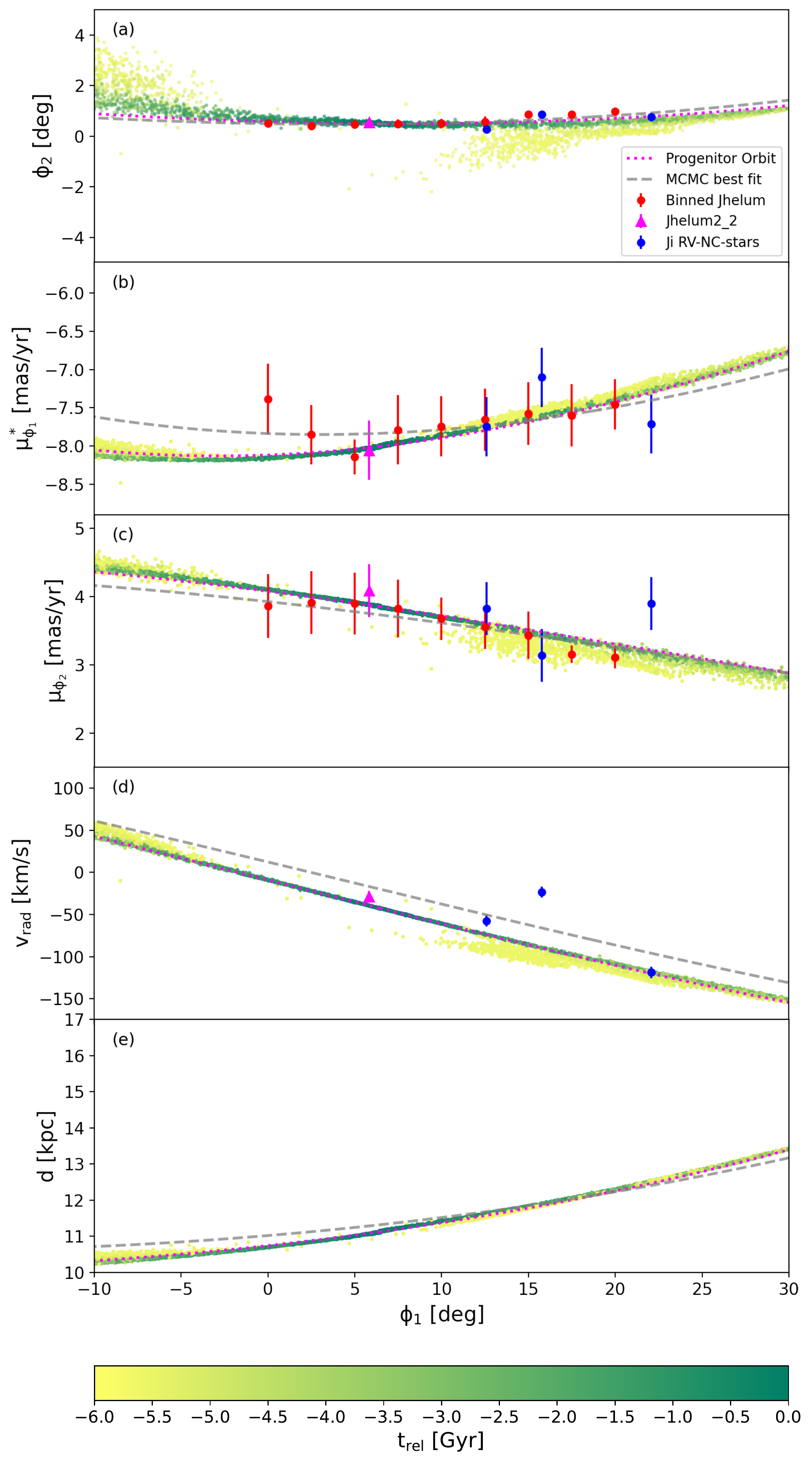}
\caption{As Fig.~\ref{fig:Sim1}, but now for Simulation 2. {The bottom component is short, dense and close to the narrow component. To the left, the leading tail shows signs of a diffuse tertiary component that is the result of a `folding'.}
\label{fig:Sim2}}
\end{figure}

\begin{figure}[htb!]
\includegraphics[width=.95\linewidth]{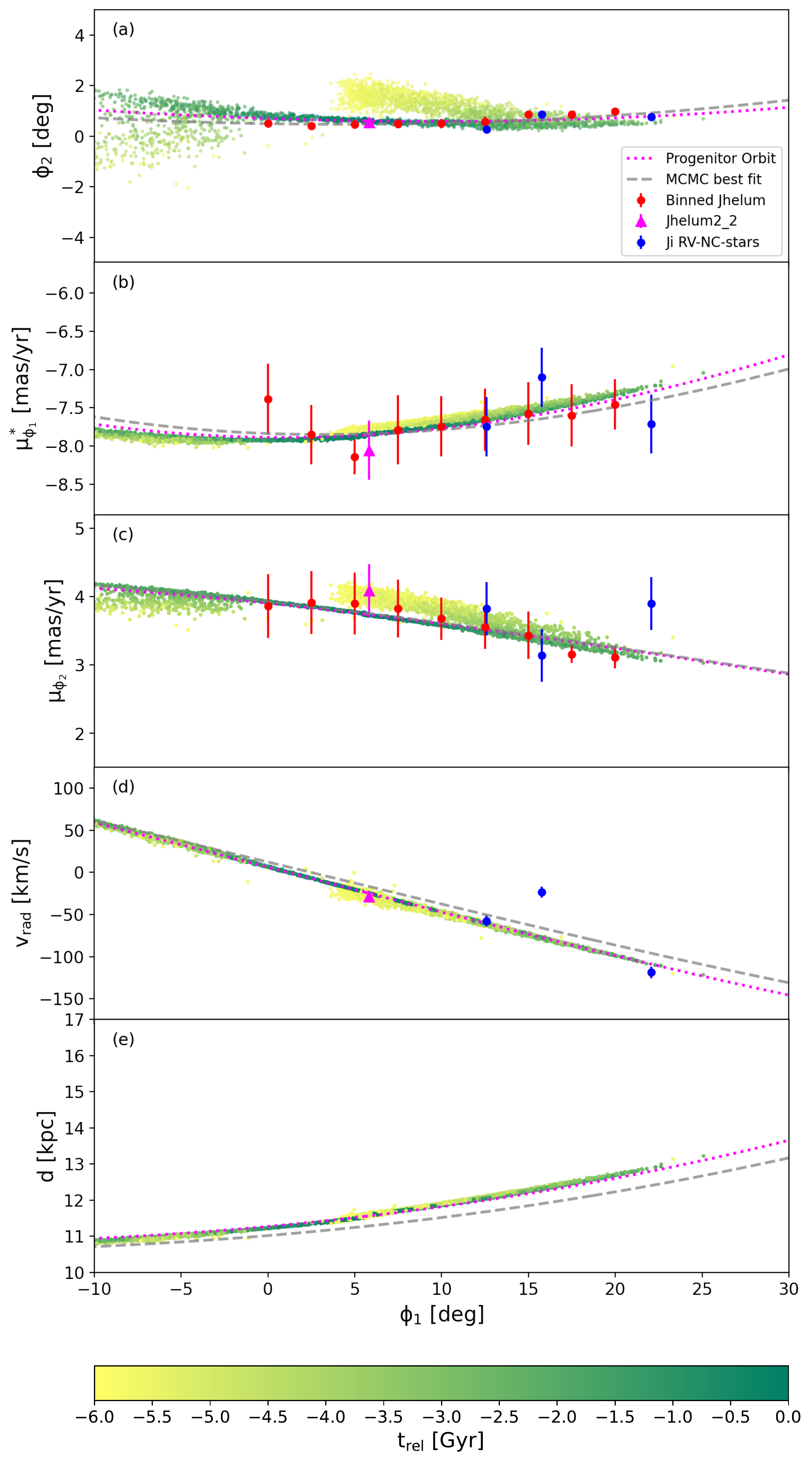}
\caption{As Fig.~\ref{fig:Sim1}, but now for initial conditions $3$. {The trailing tail has completely folded over and shows up as the component on top of the narrow component, and the leading tail shows up as a tertiary component to the bottom of the narrow component on the left of the figure. The two additional components are clearly visible in $\mu_{\phi_2}$.}
\label{fig:Sim3}}
\end{figure}

\begin{figure}[htb!]
\includegraphics[width=.95\linewidth]{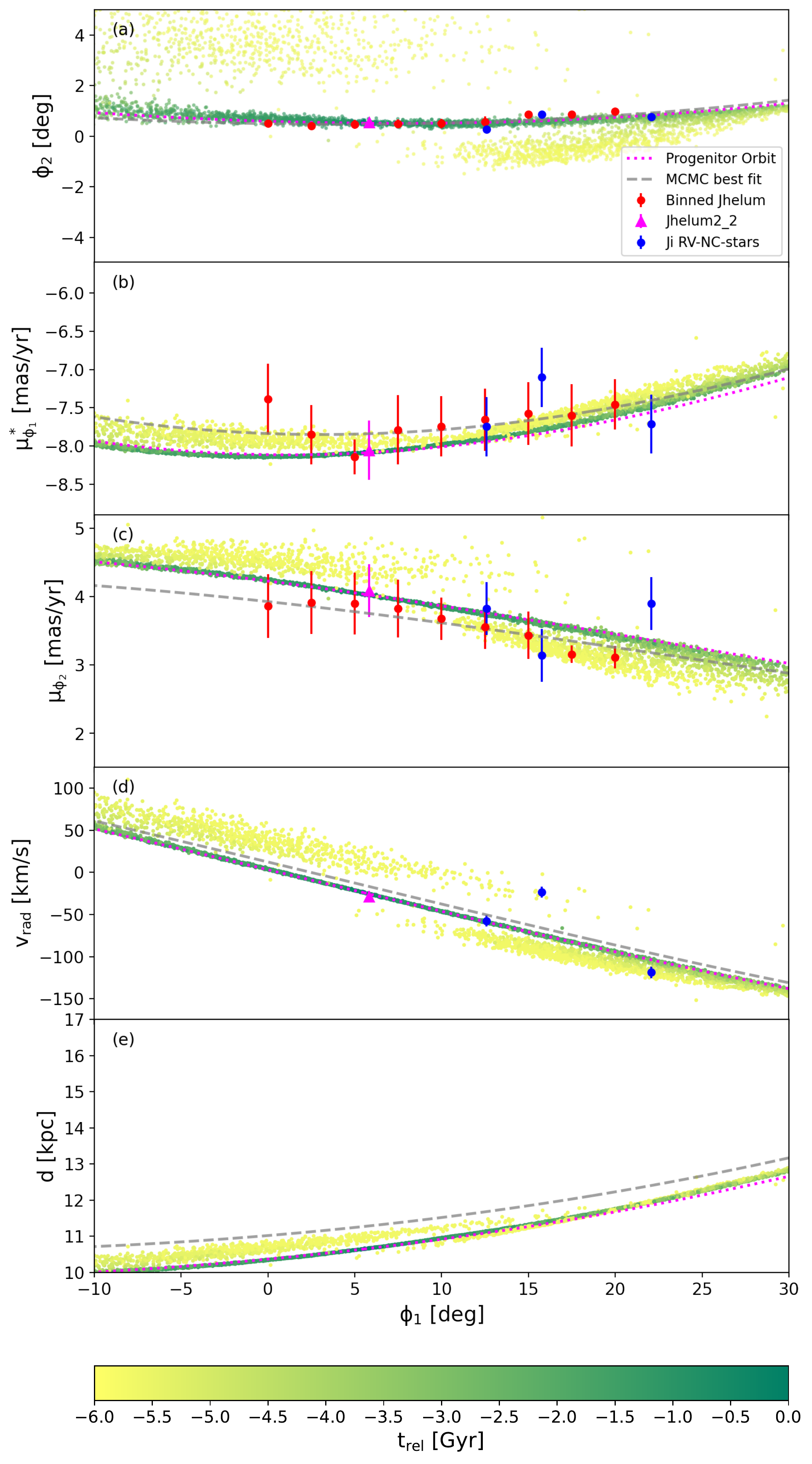}
\caption{As Fig.~\ref{fig:Sim1}, but now for initial conditions $4$. Here in Panel
    (a) we see a more diffuse secondary component in the bottom, and a tertiary component in the top, see for comparison to Figs.
    \ref{fig:Sim1}, \ref{fig:Sim2} and \ref{fig:Sim3}. This stream also has a
    very clear separation of the components in $v_{\mathrm{rad}}$, {even fitting \texttt{Jhelum1\_5}}.
\label{fig:Sim4}}
\end{figure}

  {Figures}~\ref{fig:Sim1} -- \ref{fig:Sim4} show
the present-day properties of the simulated streams using the 4 orbits described in   {Sect.}~\ref{sec:orbits_IC_Nbody}. These figures show that interactions with Sgr could certainly explain the
observed multiple components structure of the Jhelum stream 
reported in \cite{Bonaca2019Jhelum} and also seen in 
Fig.~\ref{fig:members_NC}. The simulated streams have narrow components with widths comparable to the observations. Furthermore, although the stream generally extends beyond $\phi_1=[-10,30]$~deg in our simulations, such portions of the stream are more diffuse, more
sparsely populated {or even completely `folded' over,} explaining why they may have not yet been detected. This may also apply to some of the subcomponents seen in these figures, which would perhaps not be apparent with current datasets, as their density may be too low.

A remarkable feature of all additional components is that they seem to be
formed by particles that were released from the progenitor quite early during
the formation of the stream (i.e.  { the yellow particles in   {Figs.}~\ref{fig:Sim1}--\ref{fig:Sim4}}).  This means that the interaction(s) with Sgr cause
the tail(s) of the stream to end up closer to the progenitor (located at $\phi_1=5.8$~deg) than they would if
the stream had evolved in a smooth potential.  

The diversity of the streams revealed in Figs.~\ref{fig:Sim1}--\ref{fig:Sim4}, particularly in terms of the morphology and number of subcomponents, shows that the specifics of the encounters leave different kinds of imprints on the stream. To shed light on this, we   {studied} the history of the stream and   {characterized} the interactions as described in   {Sect}.~\ref{sec:encounter:interactions}. 
To establish exactly which encounter is responsible for a specific morphological feature we   {ran} additional
simulations where we   {removed} Sgr after   {each} interaction, so that we   {could} isolate the effect of a single encounter (in practise we   {removed} Sgr when it is maximally distant from Jhelum after the interaction) and   {evolved} Jhelum in
isolation for the remainder of the time. We can thus record the impact of each
interaction before adding the next interaction to finally study the combined
effect.

The first close interactions in our simulations occur typically before $t_{\mathrm{lookback}}=-5.1$~Gyr{. R}egardless of the velocity kick imparted{, these interactions} do not correspond with any visible
substructure in the resulting stream. This is because the
stream barely started forming at this point in time in our simulations.

Interestingly, the most impactful interactions (whose properties are listed in Table \ref{tab:Intparams}) correspond to close passages of Sgr to the stream that have a differential effect. They impart the highest $\Delta v$ locally, and simultaneously impart
velocity kicks of different amplitude to different
portions of the stream. This gradient in $\Delta v$ causes
the two tails to get a different velocity kick compared to each other, and to the
progenitor. After the interaction the progenitor continues disrupting, forming
new tails. This is why we often observe three components in the case of an
impactful interaction. One   {component} corresponds to the portion of the stream formed after the interaction, while the other two   {components} are
the tails of the stream already present at the time of the interaction.

{After one impactful interaction has perturbed the original tidal tails, causing them to start precessing around the orbit of the progenitor, further close interactions with Sgr can still occur. Because the stream is now a more extended object, it is easier to create a difference in $\Delta v$ between the new stream and the two perturbed original tails. The perturbed tails are subject to a higher or lower $\Delta v$ than the particles in the newly formed tidal tails from the progenitor, which generally do not experience a gradient in $\Delta v$ in these interactions. Therefore, while these interactions will have an effect on the morphology of the stream at present time, they will generally not create additional substructure.}

The orbital phase of each interaction also has an impact on its outcome{,} in the 
sense that interactions closer to pericentre are more likely to produce a larger 
gradient in $\Delta v$ due to the contraction of the stream in this orbital phase. 
Furthermore, the geometry of the interaction itself is
also important for the present-day on-sky visibility of the resulting
substructure. We now proceed to describe in detail the characteristics of each one of the simulations shown in Figs.~\ref{fig:Sim1} -- \ref{fig:Sim4}.

\begin{table}
\caption{{Parameters of the most impactful interactions between Jhelum and Sgr}\label{tab:Intparams}}
\centering
\begin{tabular}{c|cccc}
\hline\hline
 & $b$ (\si{kpc}) & $t_{\mathrm{int}}$ (\si{Gyr}) & $v_{\mathrm{rel}}$ (\si{km.s^{-1}}) & $\Delta v_{\mathrm{max}}$ (\si{km.s^{-1}}) \\
\hline   
1 & 2.1 & -3.1 & 474.0 & 12.8 \\
2a & 2.5 & -4.3 & 619.3 & 8.4 \\
2b & 3.0 & -1.3 & 599.7 & 7.4 \\
3 & 2.2 & -2.5 & 445.7 & 14.3\\
4a & 2.5 & -4.3 & 594.8 & 8.8 \\
4b & 1.6 & -3.7 & 617.4 & 11.7 \\
\end{tabular}
\tablefoot{Interaction parameters for the closest interactions between Jhelum and
Sgr within the past $6$~Gyr for the initial conditions from Table
\ref{tab:Simprops}. The impact parameter $b$ is defined by the 100 particles of
Jhelum closest to Sgr at the interaction, $t_{\rm int}$ is the time of
this interaction, $v_{\rm rel}$ the relative velocity between the objects, and
$\Delta v_{\mathrm{max}}$ is the maximal velocity kick experienced by the
particles of Jhelum, calculated following \cite{ErkalBelokurov2015}.}
\end{table}

\subsubsection{Simulation 1}\label{sec:encounter:Sim1}

This stream, shown in Fig.~\ref{fig:Sim1}, is particularly interesting because
it shows three clear components in the $\phi_1-\phi_2$ plane,   {that is to say} on the sky. One corresponds
to the expected {narrow component}, which we also call the main component. There are
additional components above and below the main one, that could correspond to the
components seen in the data and shown in Fig.~\ref{fig:members_NC}. These additional components also
show up in the $\phi_1-\mu_{\phi_1}$ {and the $\phi_1-\mu_{\phi_2}$} plane, separated from the main component
by {up to} $\Delta\mu_{\phi_2}\sim 0.2$~\si{mas.yr^{-1}}. They are also visible in the
$\phi_1-v_{\mathrm{rad}}$ plane, but then separated by {up to} $\Delta
v_{\mathrm{rad}}\sim 30$~\si{km.s^{-1}}. 

In this simulation, {four close interactions with Sgr happen, only one of which, the one at $-3.1$~\si{Gyr}, is the main cause of the observed substructure.
This interaction imparts} a kick of $\sim{14}$~\si{km.s^{-1}} to the bottom component, $\sim{8}$~\si{km.s^{-1}} to the
main component and $\sim{4}$~\si{km.s^{-1}} to the top one, due to passing closer to the trailing tail.
The narrow component is approximately ${9}\times$ as dense as the top component, and ${12}\times$ denser than the bottom
component at the present time.

\subsubsection{Simulation 2}\label{sec:encounter:sim2}

This stream, plotted in Fig.~\ref{fig:Sim2}, is specifically interesting because
it shows only one clear secondary component, which is {2.5}$\times$ more diffuse than the main
component, {although the leading tail shows up as a diffuse folded over component at $\phi_1\sim-10$}. The {length} of the bottom component does not exactly match that reported in \cite{Bonaca2019Jhelum}, which
can be explained by the unknown position of the progenitor, which in our simulation is located far to the left in~$\phi_1$. The second component is also visible in the $\mu_{\phi_1}^*$, $\mu_{\phi_2}$
and $v_{\mathrm{rad}}$ panels of Fig.~\ref{fig:Sim2} {as a broader part of the stream.
The diffuse leading arm has a similar in origin to the observed tertiary component, which is approximately in the same location.}

The morphology found in this simulation is {mainly} due to one interaction (see Table~\ref{tab:Intparams}). The secondary component is made up from the trailing tail of the
stream, where the leftmost part consists of particles that
were released in the first pericentric passage. This encounter {at $\sim -4.3$\si{Gyr}} imparted a
velocity kick of $\sim{2}$~\si{km.s^{-1}} to the particles in the leading tail,
which should be compared to the kick of $\sim{6}$~\si{km.s^{-1}} to the particles in the main
component and $\sim{9}$~\si{km.s^{-1}} to the particles in the second component. {The other reported passage in Table~\ref{tab:Intparams} enacts a velocity kick of $\sim7.5$~\si{km.s^{-1}} to the particles in the trailing tail and the already folded over second component, $\sim6$~\si{km.s^{-1}} to the progenitor and $\sim4$~\si{km.s^{-1}} to the leading tail, which was already slightly folded over and more diffuse at this time. This reinforces the fact that not all close interactions necessarily cause substructure. Especially later interactions only enact a differential $\Delta v$ among the three components of the stream because of the already existing substructure.}

\subsubsection{Simulation 3}\label{sec:encounter:sim3}
This stream, shown in Fig.~\ref{fig:Sim3}, again shows three components, one below and one above the narrow component. {It is interesting because it matches the observed substructure quite closely when mirrored around $\phi_2=0$, and because the trailing tail is fully truncated and folded over and shows up above the narrow component}.
The bottom component is also visible in $\mu_{\phi_2}$ {as a broader part of the stream,}
while the top component is only clearly separated from the narrow component in
$\mu_{\phi_2}$, by $\Delta~\mu_{\phi_2}\sim~0.1$~\si{mas.yr^{-1}}.
The bottom component is made up of particles from the leading tail while the top one by particles from the trailing tail. These subcomponents formed after one interaction at $-2.5$~\si{Gyr},   {where Sgr} passes closest to the middle of the
leading tail and imparts a velocity kick of $\Delta v \sim {15}$~\si{km.s^{-1}} to the leading tail, $\sim {10}$~\si{km.s^{-1}} to the progenitor and $\sim
{5}$~\si{km.s^{-1}} to the trailing tail.
The narrow component is {equally dense as} the top 
component, and ${8}\times$ more dense than the bottom one.

\subsubsection{Simulation 4}\label{sec:encounter:Sim4}

The stream plotted in   {Fig.}~\ref{fig:Sim4} {has a} narrow component {that} is $\sim 10\times$ denser than the top
component, and {2}$\times$ than the bottom one. This latter
component is visible in $\mu_{\phi_2}$ and $v_{\mathrm{rad}}$,
separated by up to $\Delta \mu_{\phi_2}\sim 0.2$~\si{mas.yr^{-1}} and $\Delta
v_{\mathrm{rad}}\sim25$~\si{km.s^{-1}} from the narrow component, while the top component is separated
 in $\mu_{\phi_1}^*$, $\mu_{\phi_2}$,
$v_{\mathrm{rad}}$ and distance by up to $\Delta~\mu_{\phi_1}^*\sim~0.1$~\si{mas.yr^{-1}}, $\Delta~\mu_{\phi_2}\sim~0.4$~\si{mas.yr^{-1}}, $\Delta~v_{\mathrm{rad}}\sim~50$~\si{km.s^{-1}} and $\Delta~d\sim~0.2$~\si{kpc}.

The bottom component is made up of particles from the trailing tail, while that in the 
top consists of those from the leading tail. These components formed through two interactions, whose specifics are collected in
Table~\ref{tab:Intparams}. The first interaction causes the existing tails to
start precessing around the orbit, but this does not show up in the
$\phi_1-\phi_2$-coordinates because the velocity kick was mainly imparted in
the radial direction. The second interaction then imparts a kick that separates the tails visibly in $\phi_1-\phi_2$. In this case the closest passage of Sgr is with the middle of the leading tail,
imparting a kick of up to $\sim {14}$~\si{km.s^{-1}}, while the progenitor and
the trailing tail receive kicks of $\sim {7}$~\si{km.s^{-1}} and $\sim
{3}$~\si{km.s^{-1}} respectively.

\section{Discussion and conclusions} \label{sec:disc_&_conc}

With the new Gaia EDR3 data, we unveil new structures in the Jhelum stellar stream. This has been possible thanks to the smaller   {PM} uncertainties in EDR3
as well as to a more sophisticated selection technique. Besides the clumpy nature
of the stream and its narrow and broad components \citep[first reported in][]{Bonaca2019Jhelum}, we   {find}
evidence for a third, also broad component above the stream main track.
Additionally, we   {find} a \emph{kink}-like feature (at $\phi_1 \approx 14$~deg) in
the narrow component of the stream. {After submission of this manuscript, these findings were confirmed by \cite{Viswanathan_RPM_2022} using a reduced   {PM} halo catalogue.}

Such kinks are generally expected around the location of the progenitor, marking the transition from the leading to the
trailing arm. However, in the case of Jhelum, the leading arm (extending towards $\phi_1 < 0)$ would be expected to be above the average stream track, while the trailing (extending towards $\phi_1 > 0$)   {would be expected }to be below it, which is the opposite of what we observe. 
The {kink} is thus more likely due to a perturbation by a substructure. In fact, this has been proposed to explain the {wiggle} seen in the GD-1 stream \citep{deBoer:2020}, and the \emph{kink} feature observed in the ATLAS-Aliqua Uma stream \citep{Li2021ATLAS-AU}, which although of larger amplitude than observed in Jhelum, has   {also} been attributed to a
close interaction with Sgr. 

\begin{figure}[htb!]
\centering
\includegraphics[width=\hsize]{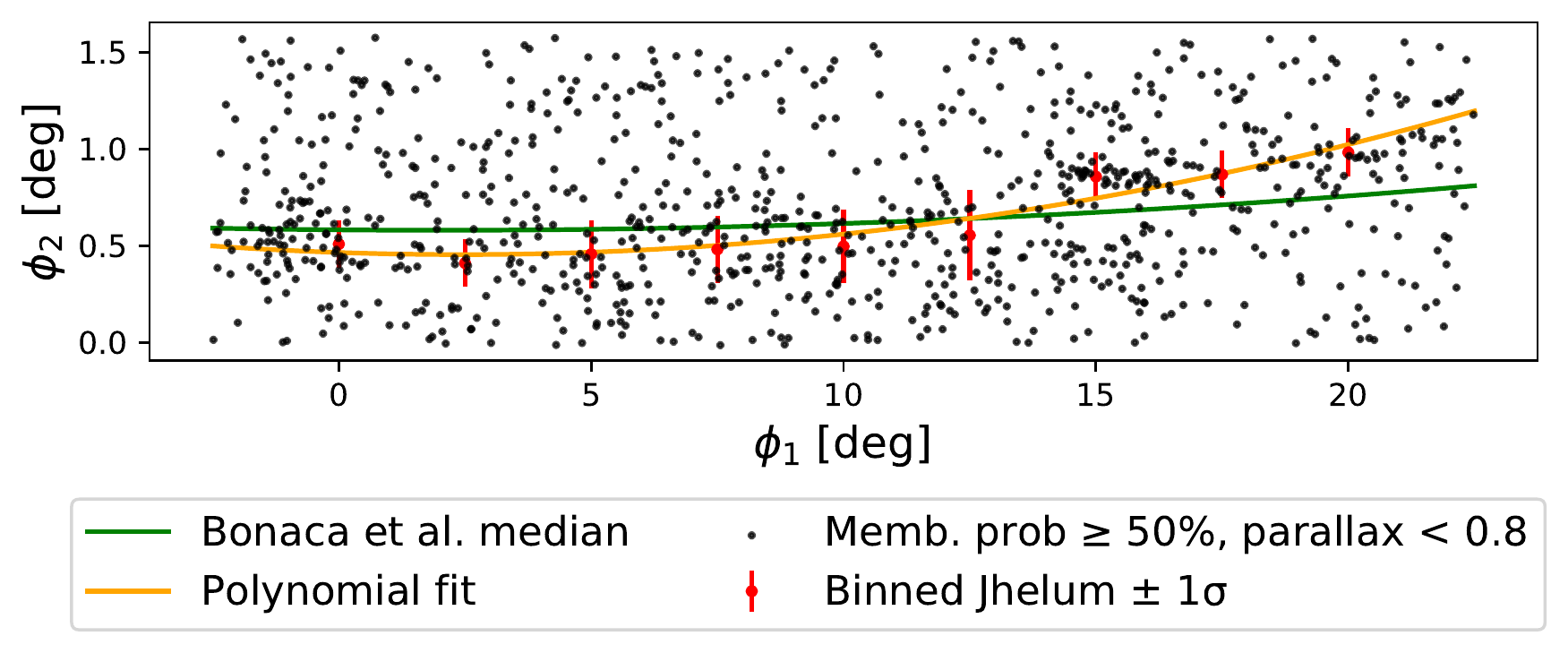}
\caption{Possible narrow component members in $(\phi_1, \phi_2)$. The binned
track with $1 \sigma$ uncertainty range are indicated in red. The polynomial
  {fitted} to this track, eq. \ref{eq:polyfit}, is shown in orange. For comparison,
the \cite{Bonaca2019Jhelum} median is overplotted in green. The updated polynomial fit
seems to follow the narrow component more closely.}
\label{fig:medianpolyfit}
\end{figure}

When characterizing Jhelum's narrow component, we find the stream track to be best described by 
the polynomial
\begin{equation}
\phi_2 = 0.00184 \phi_1^2 - 0.00891 \phi_1 + 0.466
\label{eq:polyfit}
\end{equation}
\noindent
which differs slightly from that reported by \cite{Bonaca2019Jhelum}. For
comparison, both tracks are plotted in Fig.~\ref{fig:medianpolyfit}, which shows that
the \cite{Bonaca2019Jhelum} polynomial is not curved strongly enough to follow the
binned track of the stream. To supplement the stream track data, we   {used}   {the} publicly available $S^5$ \citep{JiRVJhe} and APOGEE \citep{Sheffield2021} data, which provide radial velocity data for 9 confirmed Jhelum members. Out of these, 4 stars can be associated to Jhelum's narrow component.

We   {fitted} single orbits through the newly computed stream track and the 4 spectroscopic narrow component members
first using a standard   {MW} potential \citep{galaZenodo}. 
The best-fit orbit, which however does not fit all members of Jhelum, has a { peri}center of 8.2 kpc and an { apo}center of $23.6$~kpc. We then also explored  Galactic potentials where we allowed the characteristic parameters of the halo component to vary. In this case, the best-fit orbit has similar orbital parameters, also for different values of the internal velocity dispersion of the stream (assumed to range from 2 to 6.7~\si{\: km.s^{-1}}, but we even tested 14~\si{\: km.s^{-1}}). In all our fits, we find that the phase-space coordinates (particularly the line-of-sight velocities) of the members \texttt{Jhelum1\_5} and 
\texttt{Jhelum2\_14} are inconsistent with the track delineated by the orbits, independent of the assumed Galactic potential model. 

Although we do not expect that a single orbit will fit the exact path of the (narrow component of the) stream \citep{Sanders2013}, the discrepancies are large enough to suggest that other factors play a role in the dynamics of the stream. We therefore   {investigated} the effect of possible large perturbers, such as Sgr and the LMC. 
By integrating their orbits in a standard Galactic potential, we note that Sgr and Jhelum share roughly the same orbital plane. Moreover, Sgr and Jhelum passed each other closely multiple times 
at distances between $3 - 6$~\si{kpc} in the past $6$~\si{Gyr}. { In contrast, the LMC is currently at first infall, and hence the current close passage with the LMC at 44 kpc has been too recent for substructure to form.}
That   {Jhelum and Sgr} share approximately the same orbital plane, could possibly be explained 
if they were originally associated, as suggested by
\cite{Bonaca2021} based on their similar total energy and $z$-angular momentum. {These conclusions are robust and not dependent on the presence of a bar, the effect dynamical friction or the effect of the LMC on the orbit of Jhelum or Sgr.}

To further investigate the encounters between Jhelum and Sgr, we   {performed} N-body simulations of a dissolving, loose   {GC} on Jhelum's orbit
in the presence of Sgr using a set of four reasonable initial
conditions, consistent with the single orbit fits. We find that our N-body simulations are able to produce, at least
qualitatively, multiple components in Jhelum's stream, without fine-tuning of the
underlying potential or Sgr' orbit. {We thus   {show} that already in a relatively simple scenario we can explain qualitatively the complex present-day morphology of the Jhelum stream. In an analogous study submitted after our work, \cite{Dillamore2022} show that similar stream morphologies can be found when the effect of Sgr is considered on GD-1-like streams.}

In our simulations, we find that not only a large velocity kick locally on a portion of the stream can have a large impact, but more often, a  velocity kick gradient along the stream appears to be the main cause of the subcomponents observed in Jhelum. We also find that the geometry of the stream at the time of interaction plays a role, such that interactions at pericentre have a larger effect. 
{Furthermore, the tails need to be long enough to be observed as additional components at present time.} {The interactions causing the substructure   {are} all passages between Sgr and Jhelum within $3$~\si{kpc}, which provides a strong constraint on the orbits of both objects. From our sample of ICs for Jhelum, about 30\% of the orbits show significant substructure on-sky in the observed window, while all of the ICs produced folded or diffused tails, which did not necessarily end up to be visible in the observed window. When an interaction appears at a later time, the interaction needs to have been closer to produce an effect of a similar magnitude as an earlier interaction due to the decreasing mass of Sgr, and the fact that the additional components need time to evolve to be observable at present time.}  All our simulations match reasonably well the track followed by the narrow component, in particular the radial velocities of 3 of the 4 narrow component members. In one of our simulations the star \texttt{Jhelum1\_5} is also fitted well, although it is associated to a secondary stream component in our model.

Since the location of the progenitor {of Jhelum} is unknown, a possibility could still be that it is hiding in the region of the Galactic disk, which is situated slightly beyond $\phi_1=30$~deg, which would make the observed stream part of the leading arm. {This would however have the drawback that our formation scenario for the additional components would be unlikely.}

Additionally, in our simulations a   {GC} progenitor is able to produce all the features
currently observed in the Jhelum stream. Although 
newer $S^5$ data \citep{Li2021S5} for a larger number of stream
members indicate a velocity dispersion of $\sigma_{v_{rad}} =
13.7^{+1.2}_{-1.1}$, this estimate does not distinguish between the narrow
and broad component. In fact, we find in our simulations that Sgr is able to inflate the measured velocity dispersion of the radial velocity by a factor of up to ${4}$ compared to the unperturbed stream, with an average dispersion increase factor between 1.5 and 2, depending on which simulation and region of the stream is analysed.  { We note }however that a GC progenitor would be somewhat
inconsistent with the metallicity spread reported by \citet{Li2021S5}. Nonetheless,  a   {DG} progenitor like 
the ultra-faint dwarf (UFD) galaxy Tucana III \citep{DrlicaWagner:2015}, could produce
a stream with some of the features of Jhelum. This UFD is 
known to have associated tidal tails with a width and velocity dispersion 
comparable to what we estimate for Jhelum's narrow component \citep{Li_tuc3:2018}. 
Tucana III also has a stellar mass roughly consistent 
\citep[$\rm \sim10^3M_{\odot}$;][]{Simon:2017} with what we assume for the 
progenitor of Jhelum.

Although other possible explanations for the observed substructure in {stellar streams} have been suggested in the literature \citep[see e.g.][]{Erkal2017Pal5, Bonaca2019Jhelum,JiRVJhe,Shipp2019pmcomp,MalhanAccreted2021,Qian2022},
the scenario proposed here {for Jhelum specifically} seems natural and rather unavoidable. 
{More extensive simulations that match exactly the features seen in the stream are outside the scope of this paper, but could be used to constrain the details of the encounter as well as the Galactic potential. Such simulations could also include, for example, the reflex motion of the MW due to the infall of the LMC or even Sgr  \cite[e.g.][]{ShippLMC2021, VasilievTango2021}, as well as more complex progenitors including rotating GC \citep{Bianchini2018,Sollima2019,Erkal2017Pal5}}. That Jhelum might have been perturbed by Sgr raises attention for all
other heavy and large structures present in the   {MW} halo, such as
  {GC} and   {DGs}. We   {show} that interactions with
these heavy structures could possibly perturb streams and leave an imprint in
the form of substructure. Thus, when Sgr, or a similar structure, has
(periodic) close encounters with an object like Jhelum, we can expect other
interesting stream morphologies \citep[see also][]{2022MNRAS.510.2437E}. In the future, we would like to be able to observe a stream and discover its dynamical history from the present-day (sub)-structure. Therefore, future work should be done to {not only understand previously studied causes of substructure, but also} understand the effect of stream-subhalo interactions, specifically with heavier and larger structures like Sgr, and it should study the evolution of the stream after such
interactions. Stellar streams promise to be a valuable tool in studying the DM
subhalo population, but before we can make use of this we must first understand
all possible causes of specific stream substructure, and interactions with
objects like Sgr should be added to that list.


\begin{acknowledgements}
{We are grateful to the referee for a constructive report which led to improvements in the manuscript.} This work has made use of data from the European Space Agency (ESA) mission
{\it Gaia} (\url{https://www.cosmos.esa.int/gaia}), processed by the {\it Gaia}
Data Processing and Analysis Consortium (DPAC,
\url{https://www.cosmos.esa.int/web/gaia/dpac/consortium}). Funding for the DPAC
has been provided by national institutions, in particular the institutions
participating in the {\it Gaia} Multilateral Agreement.

This project used public archival data from DES.
Funding for the DES Projects has been provided by the DOE and NSF (USA), 
MISE (Spain), STFC (UK), HEFCE (UK), NCSA (UIUC), KICP (U. Chicago), 
CCAPP (Ohio State), MIFPA (Texas A\&M), CNPQ, FAPERJ, FINEP (Brazil), 
MINECO (Spain), DFG (Germany) and the collaborating institutions in the 
Dark Energy Survey, which are Argonne Lab, UC Santa Cruz, University of 
Cambridge, CIEMAT-Madrid, University of Chicago, University College
London, DES-Brazil Consortium, University of Edinburgh, ETH Z{\"u}rich, 
Fermilab, University of Illinois, ICE (IEEC-CSIC), IFAE Barcelona, 
Lawrence Berkeley Lab, LMU M{\"u}nchen and the associated Excellence 
Cluster Universe, University of Michigan, NOAO, University of Nottingham, 
Ohio State University, OzDES Membership Consortium, University of Pennsylvania, 
University of Portsmouth, SLAC National Lab, Stanford University, 
University of Sussex, and Texas A \& M University.
Based in part on observations at CTIO, NOAO, which is operated by 
AURA under a cooperative agreement with the NSF.

Throughout this work, we've made use of the following packages: \texttt{astropy} \citep{Astropy},
          \texttt{gala} \citep{softwarecitegala, galaZenodo}, 
          \texttt{emcee} \citep{emcee}, 
          \texttt{corner} \citep{corner}, 
          \texttt{vaex} \citep{vaex2018},
          \texttt{SciPy} \citep{2020SciPy-NMeth},
          \texttt{matplotlib} \citep{matplotlib},
          \texttt{NumPy} \citep{Numpy},
          \texttt{GADGET-4} \citep{GADGET-4},
          \texttt{AGAMA} \citep{AGAMA} and Jupyter Notebooks \citep{JupyterNotebook}.
\end{acknowledgements}

\bibliography{bibliography}
\bibliographystyle{aa} 


\begin{appendix}

\section{Comparison between a single orbit and the track defined by a stream}
\label{app:simple-Nbody}

Single orbits and stellar streams on those orbits are generally expected to be misaligned \citep[because of the spread in energies in the progenitor system, see][]{Sanders2013}. The magnitude of this effect depends on the characteristics of the orbit,  {that is to say} the regions it probes in the host potential. We therefore  {explored} in this Appendix the degree to which an orbit can be used to fit the Jhelum stream. 

\begin{figure}[hbt!]
\includegraphics[width=.95\linewidth]{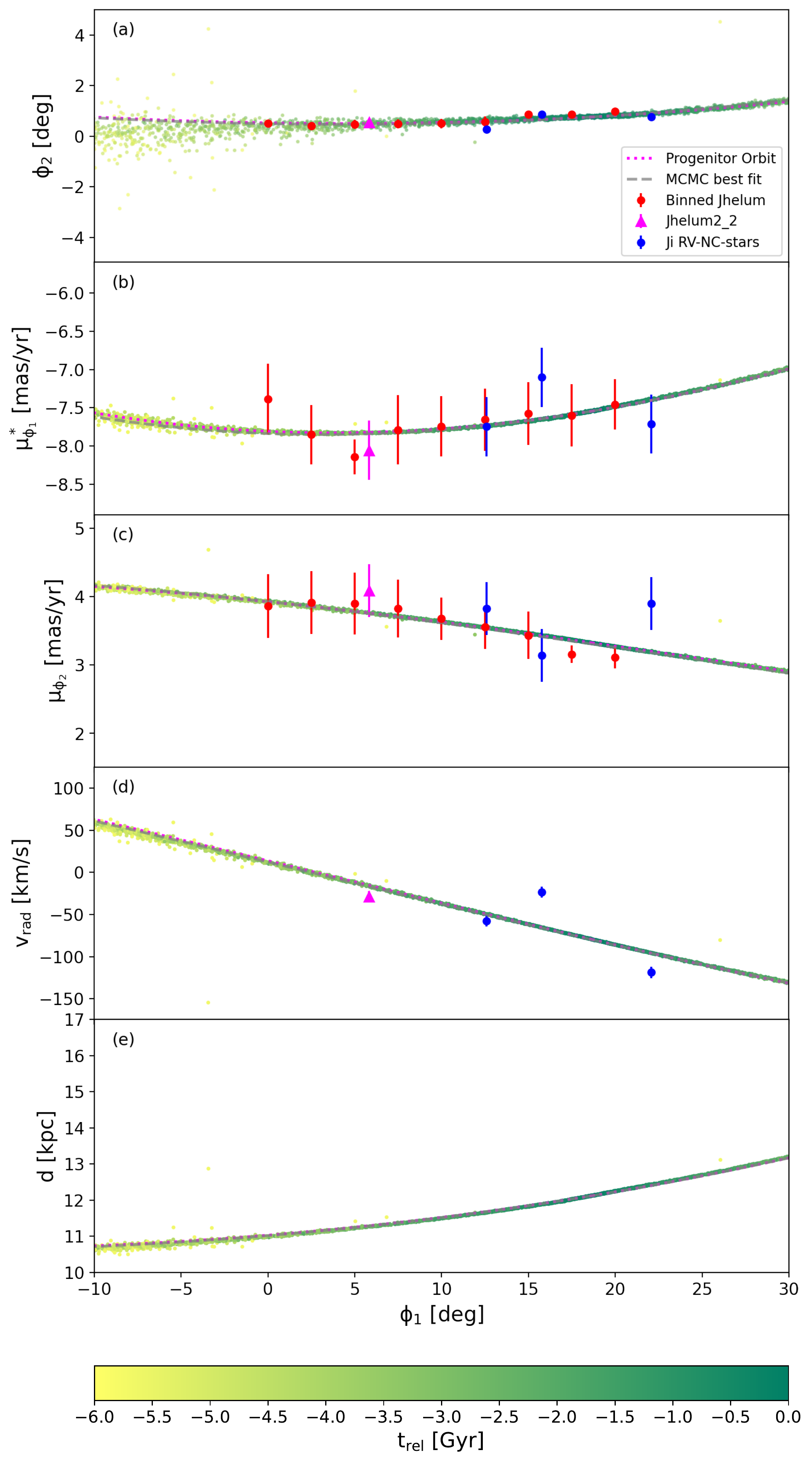}
\caption{As Fig.~\ref{fig:Sim1}, but for the same  {GC} progenitor on the best-fit orbit from Sec.~\ref{sec:results:orbitfit}. No significant differences are seen between the track of the orbit and that defined by the stream stellar particles in the N-body simulation. 
\label{fig:bestfitNbody}}
\end{figure}

We  {compared} here the track followed by the best-fit orbit found in Sec.~\ref{sec:results:orbitfit} to an N-body simulation centred on this orbit, for the  {GC} progenitor of the Jhelum stream presented in Sec.~\ref{sec:orbits_IC_Nbody}.  {Fig.}~\ref{fig:bestfitNbody} presents the results obtained. This figure shows that the differences are very small over the region covered by the data. Although a systematic trend is apparent on the sky (the top panel of Fig.~\ref{fig:bestfitNbody}) whereby the stream stellar particles for $\phi_1 < 0$~deg are slightly below the orbit, the effect is too small to lead to biased conclusions. 

\section{Neglecting the LMC} \label{app:LMCinfluence}
{ 
To ensure that the LMC can be neglected as a cause of substructure, we studied the time variation of the ratio between the forces from Sgr and the LMC on Jhelum, and compared this to the force of the MW on Jhelum to put this in context. This is shown in  {Fig.}~\ref{fig:forceratio} for Simulation 2. This behaviour is qualitatively similar for all four simulations discussed in  {Sect.}~\ref{sec:encounter:results}.

We  {calculated} the force from Sgr on Jhelum from Simulation 2, using the time-varying model for Sagittarius from Appendix \ref{app:sgrparams}. The force from the MW on Jhelum  {was} calculated using \texttt{AGAMA}. Forces from the LMC  {were} calculated with respect to the orbit of the LMC, which we find using \texttt{galpy}'s implementation of the Chandrasekhar dynamical friction force \citep{Bovy2014Potential} in the Price-Whelan potential. To determine the scale radius of the LMC we  {followed} \cite{Erkal2019}, and assuming
the present-day mass of the LMC of $1.4 \cdot 10^{11} M_{\odot}$ and a Plummer profile this gives $r_s = 4.8$ kpc and hence a half-mass radius of $r_{hm} \sim 6.3$ kpc. 

\begin{figure}[h]
    \centering
    \includegraphics[width=.95\linewidth]{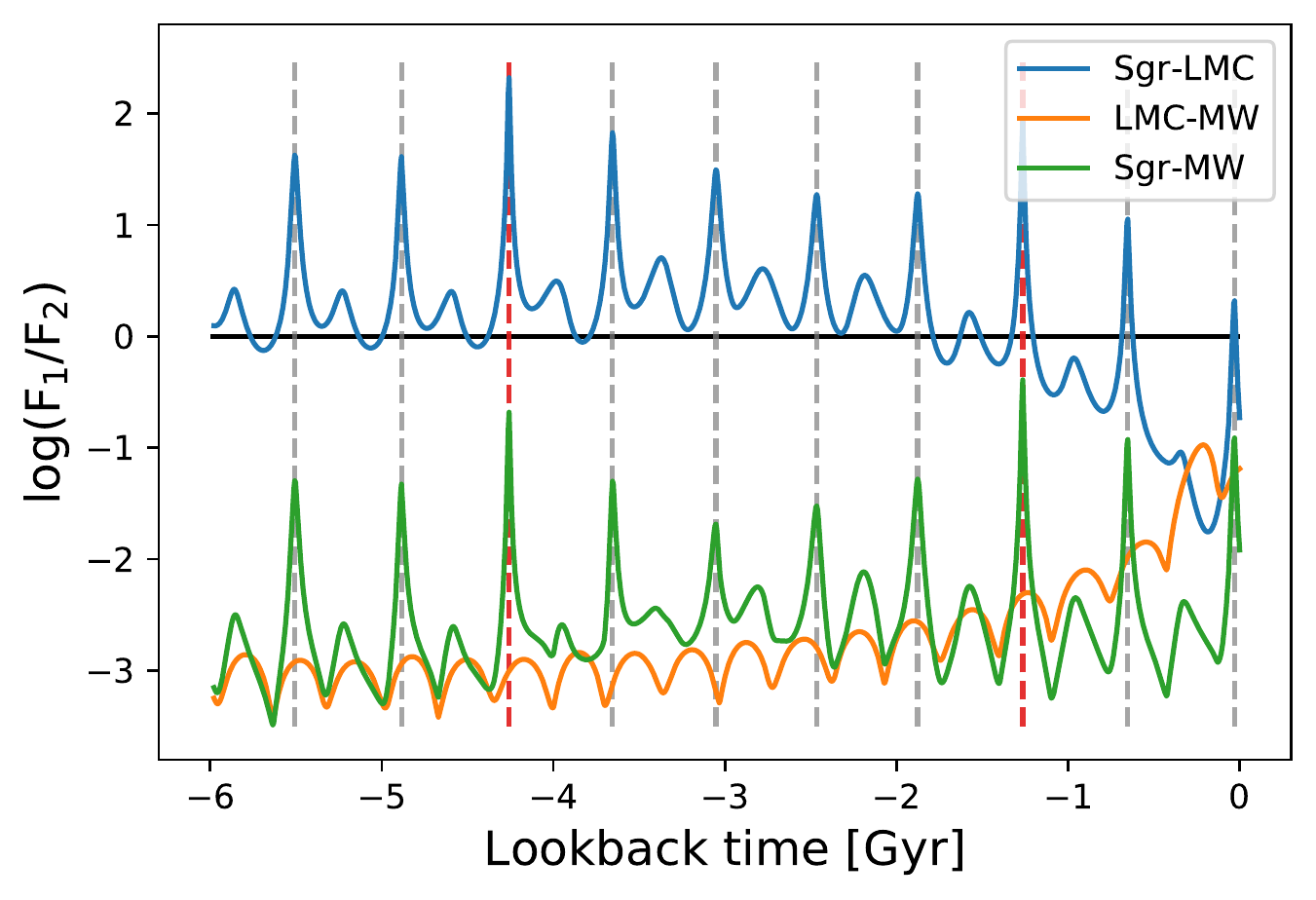}
    \caption{Logarithm of the ratio between forces of Sgr, the LMC and the  {MW} on Jhelum through time for Simulation 2. The blue line shows the ratio between the forces from the LMC and Sgr on Jhelum. The orange (green) line shows the ratio between the force from the LMC (Sgr) and the MW. The vertical dashed lines show minima in the distance between Jhelum and Sgr, with the red dashed vertical lines showing the closest, most impactful interactions from Table \ref{tab:Intparams}. The black horizontal line shows where the forces are equal.}
    \label{fig:forceratio}
\end{figure}

We see that the ratio $F_{Sgr}/F_{LMC}$ varies between $0.5-300$ before $\sim1$~\si{Gyr}, with maxima at close interactions between Jhelum and Sgr.  {We note} that during the period before $\sim1$~\si{Gyr} the force from the LMC compared to the MW is negligible (a ratio of $\sim10^{-3}$), meaning that even when $F_{Sgr}/F_{LMC} < 1$, the effect of the LMC can be neglected. On the other hand, Sgr's force reaches up to $\sim50$\% of the MW's force multiple times, specifically at close interactions between Jhelum and Sgr. At these close interactions, Sgr causes a large force gradient along the Jhelum stream which causes the final substructure.

As the LMC is at first infall, the closest passage between Jhelum and the LMC is approximately at present day.  {Figure}~\ref{fig:forceratio} shows that the LMC's force increased to up to 10\% of the MW's force during the past $\sim1$~\si{Gyr}. The close passage with the LMC causes a force gradient of up to $\sim10\%$ along the Jhelum stream (see also \cite{ShippLMC2021}). 
However, this force gradient will not cause visible substructure due to the time needed for the system to react.  }

\section{A time-varying model of Sgr} \label{app:sgrparams}

{ To model a time-varying Sagittarius
we  {followed} the procedure outlined in  {Sect.}~\ref{sec:encounter:simulation}. We set out to model Sgr as one particle with a mass and softening scale that vary with time in our simulations of the interactions with Jhelum using  \texttt{GADGET-4}. This can be done by assuming a Plummer potential for Sgr of which the mass and scale radius parameters represent the mass and softening scale of the simulated particle.

To parameterize the evolution of the  progenitor of Sgr as it disrupts in a  {MW}-like potential, we  {fitted} a Plummer model to the bound particles in each snapshot of the N-body simulation of Sgr (see  {Sect.}~\ref{sec:encounter:simulation} for details). The time evolution of the fitted scale radius and mass can be seen in  {Fig.}~\ref{fig:sgrparams} as dotted lines. Although a Plummer model does not perfectly fit the distribution of bound stars as Sgr orbits around the  {MW}, the deviation is at most $7\%$ in the mass profile.

\begin{figure}[h]
    \centering
    \includegraphics[width=0.95\linewidth]{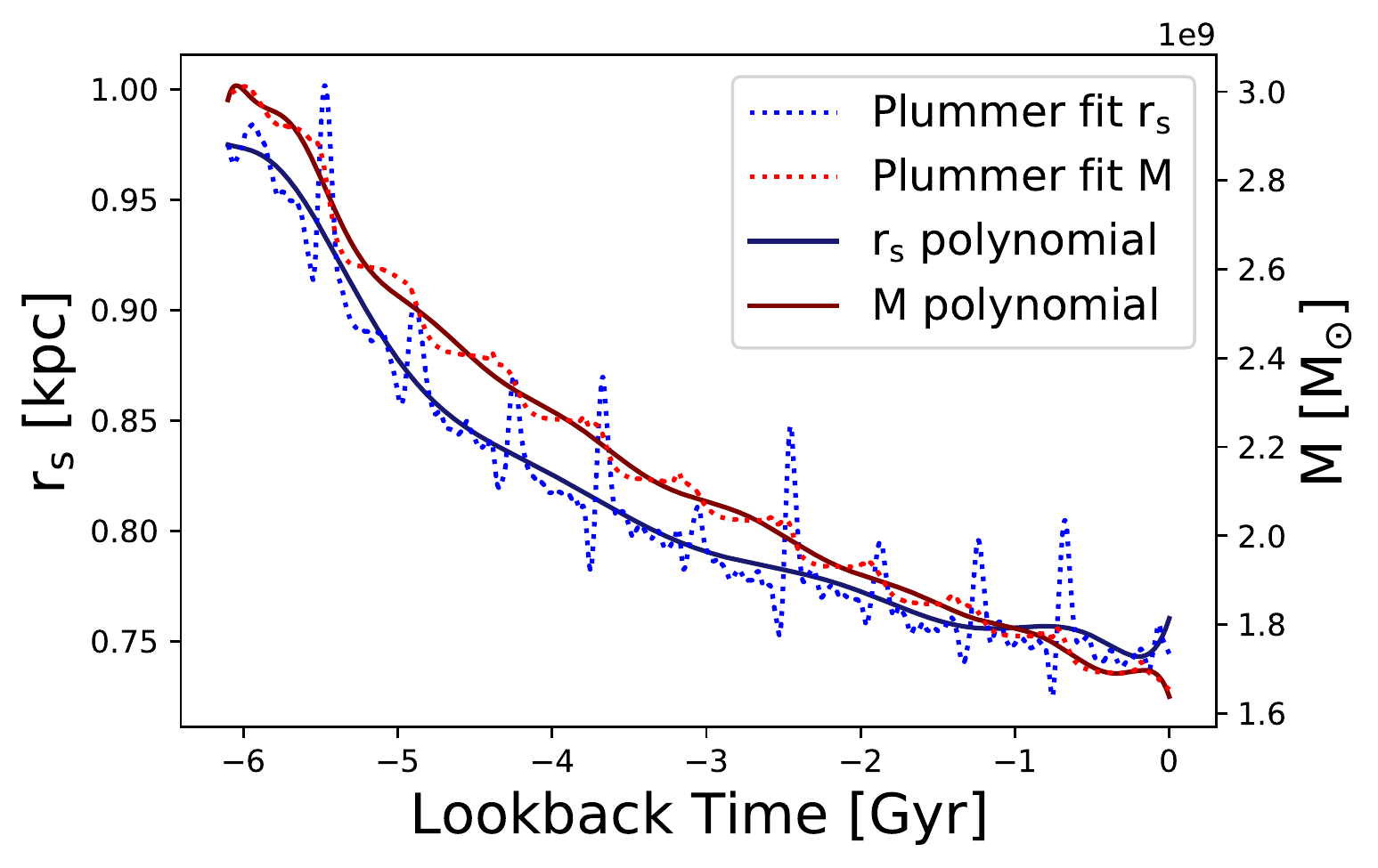}
    \caption{The smooth polynomial fit to the parameters of the Plummer model fitted to the disrupting Sgr progenitor over time. The dashed lines show the behaviour of the parameters of the Plummer fit to Sgr through the simulations. The solid lines show the smooth polynomials fitted to this behaviour.}
    \label{fig:sgrparams}
\end{figure}

After determining the values of the characteristic parameters in each snasphot, we  {fitted} a smooth polynomial model to describe their evolution in time. This is shown by the solid lines in  {Fig.}~\ref{fig:sgrparams}. This approximation is very good as the deviations in  the mass profile are up to $\sim 1\%$.

Therefore, this method of modelling a time-varying Sgr is acceptable for our general purposes of modelling the impact of a mass-varying satellite on a stellar stream. 

}

\end{appendix}

\end{document}